\documentclass[aps,prd,12pt,superscriptaddress,
amsfonts,amssymb,amsmath,byrevtex,showpacs,nofootinbib]{revtex4-1}
\pdfoutput=1
\usepackage{graphicx}
\usepackage{amssymb,amsmath,amsfonts,palatino,amsthm}
\usepackage{amssymb}
\usepackage{epstopdf}
\usepackage{color}

\begin{document}

\title{Spin Foam Vertex Amplitudes on Quantum Computer \\ - Preliminary Results}

\author{Jakub Mielczarek\footnote{jakub.mielczarek@uj.edu.pl} \\
CPT, Aix-Marseille Universit\'e, Universit\'e de Toulon, CNRS, F-13288 Marseille, France \\
Institute of Physics, Jagiellonian University, {\L}ojasiewicza 11, 30-348 Cracow, Poland}

\begin{abstract}
Vertex amplitudes are elementary contributions to the transition amplitudes in 
the spin foam models of quantum gravity. The purpose of this article is make the first 
step towards computing vertex amplitudes with the use of quantum algorithms. 
In our studies we are focused on a vertex amplitude of  3+1 D gravity, associated 
with a pentagram spin-network. Furthermore, all spin labels of the spin network are 
assumed to be equal $j=1/2$, which is crucial for the introduction of the 
\emph{intertwiner qubits}. A procedure of determining modulus squares of vertex 
amplitudes on universal quantum computers is proposed. Utility of the approach is 
tested with the use of: IBM's \emph{ibmqx4} 5-qubit quantum computer, simulator 
of quantum computer provided by the same company and QX quantum 
computer simulator. Finally, values of the vertex probability are determined employing 
both the QX and the IBM simulators with 20-qubit quantum register and compared 
with analytical predictions.
\end{abstract}

\vspace{1.5cm}

\maketitle 

\section{Introduction}

The basic objective of theories of quantum gravity is to calculate transition amplitudes 
between configurations of the gravitational field. The most straightforward approach to 
the problem is provided by the Feynman's path integral  
\begin{equation}
\langle \Psi_f | \Psi_i \rangle = \int D[g]D[\phi] e^{\frac{i}{\hslash}(S_{G}+S_{\phi})}, 
\label{PathIntegral}
\end{equation}
where $S_{G}$ and $S_{\phi}$ are the gravitational and matter actions respectively. While 
the formula (\ref{PathIntegral}) is easy to write it is not very practical for the case of continuous 
gravitational field, characterized by infinite number of degrees of freedom. One of the 
approaches to determine (\ref{PathIntegral}) utilizes discretization of the gravitational field 
associated with some cut-off scale. The expectation is that continuous limit of such discretized 
theory can be recovered at the second order phase transition \cite{Ambjorn:2012jv,Ambjorn:2011cg}. 
The essential step in this challenge is to generate different discrete space-time configurations 
(triangulations) contributing to the path integral (\ref{PathIntegral}). In Causal Dynamical 
Triangulations (CDT) \cite{Ambjorn:2012jv} which is one of the approaches to the problem, 
Markov chain of elementary moves is used to explore different triangulations between initial 
and final state. In practice, the Markov chain is implemented after performing Wick rotation 
in Eq. \ref{PathIntegral}. In the last over twenty years, the procedure has been extensively 
studied running computer simulations \cite{Bilke:1994yf}. However, in 1+1 D case analytical 
methods of generating allowed triangulations are also available. In particular, it has been 
shown that Feynman graphs of auxiliary random matrix theories generate graphs dual to 
the triangulations \cite{DiFrancesco:1993cyw}. An advantage the method is that in the 
large $N$ (color) limit of such theories symmetry factors associated with given triangulations 
can be recovered \cite{tHooft:1973alw}.  

\begin{figure}[ht!]
\centering
\includegraphics[width=8cm,angle=0]{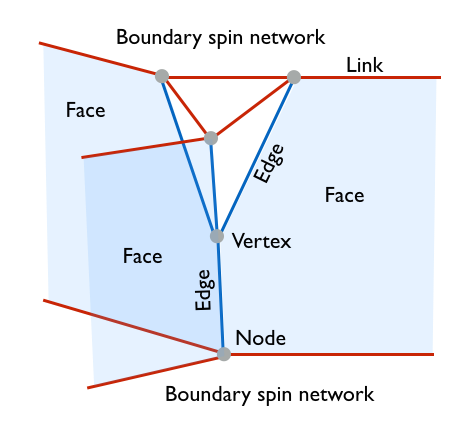}
\caption{Pictorial representation of a simple spin foam associated with 2+1 D gravity. 
The plot has been inspired by Figure 4.2 in Ref. \cite{CRFV}.}
\label{SpinFoam}
\end{figure}

Another path to the problem of determining (\ref{PathIntegral}) is provided by the Loop 
Quantum Gravity (LQG) \cite{Ashtekar:2004eh,CRFV} approach to the Planck scale physics. 
Here, discreteness of space is not due to the applied by hand cut-off but is a consequence of the 
procedure of quantization.  Accordingly, the spatial configuration of the gravitational is 
encoded in the so-called \emph{spin network} states \cite{Rovelli:1995ac}. In consequence, 
the transition amplitude (\ref{PathIntegral}) is calculated between two spin network states. 
The geometric structures (2-complexes) representing the path integral are called 
\emph{Spin Foams} \cite{Baez:1997zt,Perez:2012wv}. The elementary processes contributing to 
the spin foam amplitudes are associated with vertices of the spin foams and are called 
\emph{vertex amplitudes} \cite{Engle:2007uq, Bianchi:2012nk}. The employed terminology of 
spin networks and spin foams in clarified in Fig. \ref{SpinFoam} on example of 2+1 D gravity.

In Fig. \ref{SpinFoam} two boundary spin networks with 3-valent nodes are shown. Such 
nodes are dual to two-dimensional triangles. In this article we will focus on the 3+1 D case 
in which the spin network nodes (related with non-vanishing volumes) are 4-valent. 
The nodes are dual to tetrahedra (3-simplex). In the example presented in Fig. \ref{SpinFoam}
the four edges of the 2-complex meet at the vertex. However,  in the 3+1 D case the valence 
of the vertex is higher and equal 5. For the purpose of this article it is crucial to note that such 
5-valent vertices can be enclosed by a boundary represented by a spin network containing five 
nodes. Each of the node is placed on one of the five edges entering the vertex. The boundary 
has topology of a three-sphere, $S^3$. This is higher dimensional extension of the 2+1 D case, 
where a vertex can be enclosed by the two-sphere, $S^2$. 

In analogy to the random matrix theories in case of the 2D triangulations, the spin foams (2-coplexes) 
can also be obtained as Feynman diagrams of some auxiliary field theory. Namely, the so-called
Group Field Theories (GFTs) have been introduced to generate structure of vertices and edges 
associated with spin foams \cite{Oriti:2006se,Freidel:2005qe,Krajewski:2012aw}. In particular, 
the 3+1 D theory with 5-valent vertices requires GFT with five-order interaction terms, known as 
Ooguri's model \cite{Ooguri:1992eb}.  There has recently been made a great progress in the field 
of GFTs with many interesting results (see e.g. \cite{Gielen:2013kla,Benedetti:2014qsa}).  

The aim of this article is to investigate a possibility of employing universal quantum computers 
to compute vertex amplitudes of 3+1 D spin foams. The idea has been suggested in Ref. 
\cite{Li:2017gvt}, however, not investigated there. Here, we make the first attempt to materialize 
this concept. In our studies, we consider a special case of spin networks with spin labels 
corresponding to fundamental representations of the $SU(2)$ group, for which \emph{intertwiner 
qubits} \cite{Feller:2015yta,Li:2017gvt,Mielczarek:2018ttq} can be introduced. The qubits will be 
implemented on IBM Q 5-qubit quantum computer (\emph{ibmqx4}) as well as with the use of 
 quantum computer simulator provided by the same company \cite{IBM}. In the case 
of real 5-qubit quantum computer the qubits are physically realized as superconducting circuits 
\cite{You} operating at millikelvin temperatures. Furthermore, the QX quantum computer 
simulator \cite{QX}, available on the Quantum Inspire \cite{QuantInsp} platform, will be employed.   

The studies contribute to our broader research program focused on exploring the possibility of 
simulating Planck scale physics with the use of quantum computers. The research is in the spirit 
of the original Feynman's idea \cite{Feynman:1981tf} of performing the so-called \emph{exact 
simulations} of quantum systems with the use of quantum information processing devices. 
In our previous articles \cite{Mielczarek:2018ttq,Mielczarek:2018nnd} we have preliminary 
explored possibility of utilizing Adiabatic Quantum Computers \cite{AQG} to simulate quantum 
gravitational systems. Here, we are making first steps towards the application of Universal 
Quantum Computers \cite{Deutsch:1995dw,Ekert}. 

\section{Intertwiner qubit}

The basic question a skeptic can ask is why it is worth considering quantum computers to study 
Planck scale physics at all? Can't we just do it employing classical supercomputers as in the case 
of CDT approach to quantum gravity? Let me answer to this questions by giving two arguments. 
The first concerns the huge dimensionality of a Hilbert space for many-body quantum system. 
For a single spin-1/2 (qubit) Hilbert space $\mathcal{H}_{1/2}=\text{span}\{|0\rangle,|1\rangle \}$ 
the dimension is equal 2. However, considering $N$ such spins (qubits) the resulting Hilbert 
space is a tensor product of N copies of the qubit Hilbert space. The dimension of such space 
grows exponentially with N:
\begin{equation}
{\rm dim} (\underbrace{\mathcal{H}_{1/2}\otimes  \mathcal{H}_{1/2}\otimes 
\dots \otimes \mathcal{H}_{1/2}}_{N})=2^N. 
\end{equation}
This exponential behavior is the main obstacle behind simulating quantum systems on 
classical computers. With the present most powerful classical supercomputers we can simulate 
quantum systems with $N=64$ at most \cite{Chen}. The difficulty is due to the fact that 
quantum operators acting on $2^N$ dimensional Hilbert space are represented by $2^N\times2^N$ 
matrices. Operating with such  matrices for $N>50$ is challenging to the currently available 
supercomputers. On the other hand, such companies as IBM or Rigetti Computing are developing 
quantum chips with $N>100$ and certain topologies of couplings between the qubits. Possibility 
of simulating quantum systems which are unattainable to classical supercomputers may, 
therefore, emerge in the coming decade leading to the so-called \emph{quantum supremacy} 
\cite{Biamonte}.  See Appendix A for more detailed discussion of the state of the 
art of the quantum computing technologies and prospects for the near future. The second argument 
concerns \emph{quantum speed-up} leading to reduction of computational complexity of some 
classical problems. Such possibility is provided by certain quantum algorithms (e.g. Deutsch, 
Grover, Shor,...) thanks to the so-called \emph{quantum parallelism}.  For more information
on quantum algorithms please see Appendix B, where elementary introduction to quantum 
computing can be found.

Taking the above arguments into account we are convinced that it is justified to explore the 
possibility of simulating quantum gravitational physics on quantum computers. The fundamental 
question is, however, whether gravitational degrees of freedom can be expressed with qubits, 
which are used in the current implementations of quantum computers\footnote{In general, 
quantum variables associated with higher dimensional Hilbert spaces may be considered.}? 
Fortunately, it has recently been shown that at least in Loop Quantum Gravity approach to 
quantum gravity notion of qubit degrees of freedom can be introduced and is associated with the 
intertwiner space of a certain class of spin networks (see Refs. \cite{Feller:2015yta,Li:2017gvt,
Mielczarek:2018ttq,Mielczarek:2018nnd}). 

\begin{figure}[ht!]
\centering
\includegraphics[width=6cm,angle=0]{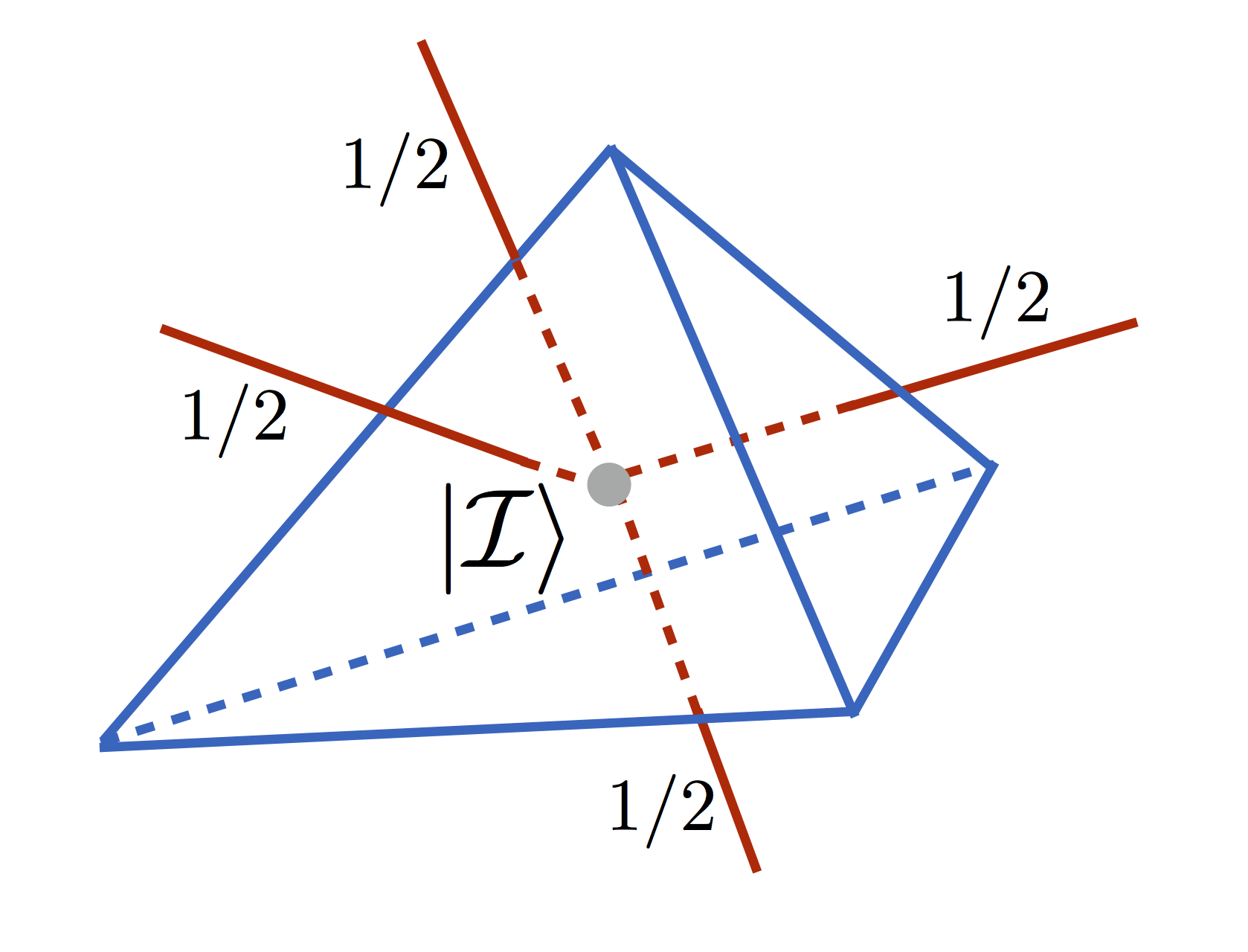}
\caption{A single 4-valent node together with the entering links with spin labels $j=1/2$. 
The intertwiner qubit $|\mathcal{I}\rangle$ is a degree of freedom defined at the node. 
The node is dual to the tetrahedron (3-symplex) as represented in the picture.}
\label{Node}
\end{figure}

Let us briefly explain it. Namely, nodes of the spin networks are where Hilbert spaces associated 
with the links meet. The gauge invariance (enforced by the Gauss constraint) implies that the total 
spin at the node has to be equal zero. The 4-valent nodes are of special interest since they are 
associated with the non-vanishing eigenvalues of the volume operator (see e.g. Ref.  \cite{CRFV}). 
As already mentioned in Introduction, in the picture of discrete geometry, the 4-valent nodes are 
dual to tetrahedra. The class of spin networks that we are focused on here are those with links of 
the spin networks labelled by fundamental representations of the $SU(2)$ group (i.e. the spin labels 
are equal $j=1/2$) and the nodes are 4-valent. For such spin networks the Hilbert spaces at the 
nodes are given by the following tensor products: 
\begin{eqnarray}
\mathcal{H}_{1/2}\otimes \mathcal{H}_{1/2}\otimes \mathcal{H}_{1/2}\otimes \mathcal{H}_{1/2} 
= 2\mathcal{H}_{0} \oplus 3 \mathcal{H}_{1}  \oplus \mathcal{H}_2. 
\label{tensorproduct}
\end{eqnarray}
There Gauss constraint implies that only singlet configurations ($\mathcal{H}_{0}$) are allowed.  Because 
there are two copies of the spin-zero configurations in the tensor product (\ref{tensorproduct}), 
the so-called intertwiner Hilbert space is two-dimensional:
\begin{equation}
{\rm dim\ Inv} (\mathcal{H}_{1/2}\otimes \mathcal{H}_{1/2}\otimes \mathcal{H}_{1/2}\otimes 
\mathcal{H}_{1/2})=2. 
\end{equation}
We associate the two-dimensional invariant subspace with the \emph{intertwiner qubit} 
$|\mathcal{I}\rangle \in \mathcal{H}_{0}\oplus\mathcal{H}_{0}$. The 4-valent node (at which the 
intertwiner qubit is defined) together with the entering links is dual to the tetrahedron in a way shown 
in Fig. \ref{Node}.

The two basis states of the intertwiner qubit $|\mathcal{I}\rangle$ are basically the two 
singlets we can obtain for a system of four spins $1/2$. The basis states can be expressed 
composing familiar singlets and triplet states for two spin-1/2 particles:  
\begin{align}
|S\rangle &= \frac{1}{\sqrt{2}} \left(|01\rangle-|10\rangle \right), \\
|T_{+}\rangle &= |00\rangle, \\
|T_{0}\rangle &= \frac{1}{\sqrt{2}} \left(|01\rangle+|10\rangle \right), \\
|T_{-}\rangle &= |11\rangle.
\end{align}
Namely, in the $s$-channel (which is one of the possible superpositions) the intertwiner qubit 
basis states can be expressed as follows:
\begin{align}
|0_s\rangle&= |S\rangle \otimes |S\rangle, \\
|1_s\rangle&= \frac{1}{\sqrt{3}} \left( |T_{+}\rangle \otimes |T_{-}\rangle+
|T_{-}\rangle \otimes |T_{+}\rangle-|T_{0}\rangle \otimes |T_{0}\rangle   \right).
\end{align}

The $|0_s\rangle$ state is simply a tensor product of two singlets for two spin-1/2 particles, while 
the state $|1_s\rangle$ does not have such simple product structure. The states $|0_s\rangle$ and 
$|1_s\rangle$ form an orthonormal basis of the intertwiner qubit. Worth stressing is that other 
bases being linear compositions of $|0_s\rangle$ and $|0_s\rangle$ might be considered.
In particular, the eigenbasis of the volume operator turns out to be useful (see Ref. \cite{Mielczarek:2018nnd}). 
Here we stick to the $s$-channel basis $\{|0_s\rangle,|1_s\rangle\}$ in which a general intertwiner 
state (neglecting the total phase) can be expressed as 
\begin{equation}
| \mathcal{I} \rangle = \cos(\theta/2) |0_s\rangle+e^{i\phi} \sin(\theta/2)|1_s\rangle, \label{IntertwinerState}
\end{equation} 
where $\theta \in [0,\pi]$ and $\phi \in [0, 2\pi)$ are angles parametrizing the Bloch sphere. 

In the context of quantum computations it is crucial to define a quantum algorithm 
(a unitary operation $ \hat{U}_{\mathcal{I}}$ acting on the input state) which will allows 
us to create the intertwiner state (\ref{IntertwinerState}) from the input state $|0000\rangle$, i.e. 
\begin{equation}
| \mathcal{I} \rangle = \hat{U}_{\mathcal{I}}|0000\rangle.  
\end{equation}
The general construction of the operator $\hat{U}_{\mathcal{I}}$ can be performed 
applying the procedure introduced in Ref. \cite{Long}, and will be discussed in a 
sequel to this article \cite{Mielczarek2019}. Here, for the purpose of illustration of the 
method of computing vertex amplitude we will focus on the special case of the 
intertwiner states being the first basis state: $| \mathcal{I} \rangle =  |0_s\rangle = |S\rangle 
\otimes |S\rangle$. The contributing two-particle singlet states can easily be generated 
as a sequence of elementary gates used to construct quantum circuits (see also Appendix B):
\begin{equation}
|S\rangle = \widehat{\text{CNOT}}(\hat{H}\otimes \hat{\mathbf{I}})(\hat{X}\otimes\hat{X})|00\rangle.      
\end{equation}
Here, the  $\hat{X}$ is the so-called \emph{bit-flip} (NOT) operator (Pauli $\sigma_x$ matrix) which
transforms $|0\rangle$ into $|1\rangle$ and $|1\rangle$ into $|0\rangle$ (i.e. $\hat{X}|0\rangle
=|1\rangle$ and $\hat{X}|1\rangle =|0\rangle$). The $\hat{H}$ is the Hadamard operator 
defined as $\hat{H}|0\rangle =\frac{1}{\sqrt{2}}\left(|0\rangle +|1\rangle\right)$ and
$\hat{H}|1\rangle =\frac{1}{\sqrt{2}}\left(|0\rangle -|1\rangle\right)$. Finally, the 
$\widehat{\text{CNOT}}$ is the Controlled NOT 2-qubit gate defined as 
$\widehat{\text{CNOT}}(|a\rangle \otimes|b\rangle) =|a\rangle \otimes|a\oplus b\rangle$, where 
$a,b \in\{ 0,1\}$ and $\oplus$ is the XOR (exclusive or) logical operation, such that
$0\oplus 0 = 0$, $0\oplus 1 = 1$, $1\oplus 0 = 1$ and $1\oplus 1 = 0$. In consequence, 
the $|0_s\rangle$ basis state can be expressed as follows:
\begin{align}
|0_s\rangle &= (\widehat{\text{CNOT}}\otimes \widehat{\text{CNOT}})
(\hat{H}\otimes \hat{\mathbf{I}} \otimes\hat{H}\otimes \hat{\mathbf{I}})(\hat{X}\otimes\hat{X}\otimes\hat{X}\otimes\hat{X})
|0000\rangle \nonumber \\
 &=    \frac{1}{2}\left(|0101\rangle+|1010\rangle -|0110\rangle -|1001\rangle\right). \label{0sstate}
\end{align}

\begin{figure}[ht!]
\centering
\includegraphics[width=16cm,angle=0]{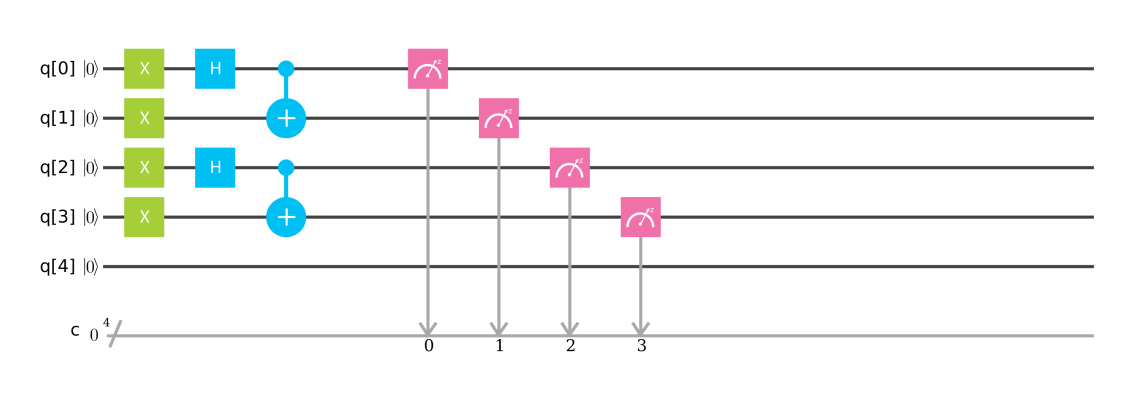}
\caption{Quantum circuit used to generate $|0_s\rangle$ state from the initial state $|0000\rangle$. 
The (green) boxes with letter X represent the bit-flip gates, the (blue) boxes with letter H represents 
the Hadamard gates, while the next operations from the left are the CNOT  2-qubit gates. Finally, 
the (pink) boxes on the right represent measurements performed at every of the four involved qubits.}
\label{Intertwiner0s}
\end{figure}

In Fig. \ref{Intertwiner0s} a quantum circuit generating (and measuring) the intertwiner state $|0_s\rangle$ 
has been presented. 
The final state can be written as a superposition of 16 basis states in the product 
space of four qubit Hilbert spaces:
\begin{equation}
|\Psi\rangle =\sum_{ijkl\in \{ 0,1\}} a_{ijkl} |ijkl  \rangle, \label{PsiMeasured}
\end{equation}
where the normalization condition implies that $\sum_{ijkl\in \{ 0,1\}} |a_{ijkl}|^2=1$. 

We have executed the quantum algorithm (\ref{0sstate}) with the use of both the IBM simulator 
of quantum computer and the real IBM Q 5-qubit quantum chip ibmqx4.  In both cases 
the algorithm has been executed 1024 times. Moreover, the algorithm (\ref{0sstate}) has also been 
executed (1024 times) on the QX quantum computer simulator. Results of the measurements 
of probabilities $P(i)=|a_i|^2$ are summarized in Table \ref{Table1}.  

\begin{table}[h!]
\begin{tabular}{|c|c|c|c|c|c|}
\hline
No. &\  Probability  \   &\ Theory \ &\ IBM simulator\  & \ IBM Q ibmqx4 \  & QX simulator  \\ \hline
1   & $|a_{0000}|^2$ &  0      & 0        & 0.014 & 0 \\
2   & $|a_{0001}|^2$ &  0      & 0        & 0.058 & 0 \\
3   & $|a_{0010}|^2$ &  0      & 0        & 0.050 & 0 \\
4   & $|a_{0011}|^2$ &  0      & 0        & 0.004 & 0 \\
5   & $|a_{0100}|^2$ &  0      & 0        & 0.023 & 0 \\
6   & $|a_{0101}|^2$ &  0.25 & 0.264 & 0.109 & 0.252 \\
7   & $|a_{0110}|^2$ &  0.25 & 0.232 & 0.091 & 0.241  \\
8   & $|a_{0111}|^2$ &  0      & 0        & 0.009 &  0 \\
9   & $|a_{1000}|^2$ & 0      & 0         & 0.034 & 0 \\
10 & $|a_{1001}|^2$ & 0.25 & 0.248  & 0.159 & 0.230 \\
11 & $|a_{1010}|^2$ & 0.25 & 0.256  & 0.158 & 0.276 \\
12 & $|a_{1011}|^2$ & 0      & 0        & 0.012 & 0 \\
13 & $|a_{1100}|^2$ & 0      & 0        & 0.034 & 0 \\
14 & $|a_{1101}|^2$ & 0      & 0        & 0.132 & 0 \\
15 & $|a_{1110}|^2$ & 0      & 0        & 0.110 &  0 \\
16 & $|a_{1111}|^2$ & 0      & 0        & 0.003 & 0 \\
\hline
\end{tabular}
\caption{Results of measurements of $P(i)=|a_i|^2$ for the quantum circuit 
presented in Fig. \ref{Intertwiner0s}.}
\label{Table1}
\end{table}

Clearly, the results obtained from the simulator matches well with the values predicted 
in Eq. \ref{0sstate}.  Increasing the number of shots an accuracy of the results 
can be improved. In fact, because the number of shots in a single round was limited (either 
to $8192$ in case of the IBM simulator or to $1024$ in case of the QX simulator)  the 
computational rounds had to be repeated in order to achieve better convergence to 
theoretical predictions. Furthermore, the results were also verified with use of two other 
publicly available quantum simulators of quantum circuits, i.e. Quirk \cite{Quirk} and 
Q-Kit \cite{QKit}.

On the other hand, the errors of the \emph{ibmqx4} quantum processor are more 
significant, leading even to $10\%$ contribution from the undesired states, such as 
$|1101\rangle$ and $|1110\rangle$. The errors have two main sources. The first are 
instrumental errors associated with both uncertainty of gates and the uncertainty of 
readouts. For the IBM Q \emph{ibmqx4} quantum processor the single-qubit gate 
errors are at the level of $0.001$ and the errors of readouts are reaching even 
$0.086$ for some of the qubits. The two-qubit gates are less accurate than the single 
quibit gates, with the errors approximately equal to $0.035$. The concrete values for 
every qubit and pairs of qubits are provided via the IBM website \cite{IBM}. The second 
source of error is due to statistical nature of quantum mechanics and the limited number 
of measurements. For a single qubit, the problem of estimating corresponding error is 
equivalent to the 1D random walk, which leads to uncertainty of the estimation of 
probability equal $\frac{s}{n}=\sqrt{\frac{p(1-p)}{n}}$, where $n$ is the number of 
measurements and $p$ is a probability of one of the two basis states. As an example, 
for $n=1024$ and $p=1/2$ we obtain $\frac{s}{n} \approx 0.016$. In the considered 
case of $16$ basis states the uncertainty is expected to be lower roughly by 
the factor $0.5$ \footnote{In order to prove it let us consider an asymmetric 1D random 
walk with probabilities $p=\frac{1}{16}$ (one of the basis states) and $q=1-p =\frac{15}{16}$ 
(rest of the 15 basis sates), for which $\frac{s}{n}=\sqrt{ \frac{1}{16} \frac{15}{16}\frac{1}{n}} 
\approx \frac{1}{4\sqrt{n}}$.}, leading to an approximate error equal $0.008$ (the 
value is smaller because the average number of counts per basis states decreased).  
Summing up both the instrumental error and the uncertainty of measurement, we may 
estimate the cumulative uncertainty to be at the level of $\sim 15\%$, which is in agreement 
with the experimental data. With the current setup, the errors can be slightly reduced 
by increasing the number of measurements and by optimization of the quantum circuit, 
e.g. by placing (less noisy) single-qubit gates after (more noisy) two-qubit gates. In the 
further studies, the circuits should be also equipped with quantum error correction 
algorithms. This will, however, require additional qubits to be involved. Moreover, 
reduction of the instrumental error will be a crucial challenge for the future utility of the 
quantum processors. 

Let us end this section with quantitative comparison of the results from the Table \ref{Table1} 
with the use of classical Fidelity (Bhattacharyya distance) $F(q,p) : = \sum_{i} \sqrt{p_i q_i}$, 
where $\{ p_i\}$ and $\{ q_i\}$ are two sets of probabilities. Comparison of the theoretical values with 
the results obtained from the IBM Simulator gives us $F \approx 99,9 \%$. Furthermore, comparing 
the theoretical values with the results of IBM Q ibmqx4 quantum computer we find $F \approx 71,4 \%$.  
However, worth keeping in mind is that the employed Fidelity function concerns classical probabilities 
and further analysis of the quantum state obtained from the quantum computer should include also 
analysis of the quantum Fidelity  $F(\hat{\rho}_1,\hat{\rho}_2) := \text{tr} \sqrt{ \sqrt{\hat{\rho}_1}
\hat{\rho}_2\sqrt{\hat{\rho}_1} }$, where $\hat{\rho}_1$ and $\hat{\rho}_2$ are density matrices of 
the compared states \cite{Nielsen}.  For this purpose (i.e. reconstruction of the density matrix) full 
tomography of the obtained quantum state has to be performed.

\section{Vertex amplitude}

Gravity is a theory of constraints. Specifically, in LQG three types of constraints are involved. 
The first is the mentioned Gauss constraint, which has already been imposed at the stage of 
constructing spin networks states. The second is the spatial diffeomorphism constraint
which is satisfied by introducing equivalence relation between all spin-networks characterized 
by the same topology. The third is the so-called scalar or Hamiltonian constraint, which encodes 
temporal dynamics and is the most difficult to satisfy. In quantum theory, this constraint 
takes a form of an operator. Let us denote this operator as $\hat{C}$. Following the Dirac 
procedure for constrained quantum systems, the physical states are those belonging to the 
kernel of the constraints, i.e. $\hat{C} | \Psi \rangle = 0$. Due to the complicated form of the 
gravitational scalar constraint (see e.g. \cite{Ashtekar:2004eh}), finding the physical states 
is in general a difficult task. However, for certain simplified scalar constraints, such as for the 
symmetry reduced cosmological models, the physical states are possible to extract. Furthermore, 
it has recently been proposed in Ref. \cite{Mielczarek:2018ttq} that the problem of solving 
simple constraints can be implemented on Adiabatic Quantum Computers.  

Another approach to the problem of constraints is to consider a projection operator 
\begin{equation}
\hat{P} := \lim_{T\rightarrow \infty} \frac{1}{2T} \int_{-T}^{T} d\tau e^{i \tau \hat{C}}, 
\label{projection} 
\end{equation}
which projects kinematical states onto physical subspace. In particular, the formula
(\ref{projection}) is valid for $\hat{C}$ characterized by discrete spectrum of eigenvalues.  
Specifically, the projection operator (\ref{projection}) can be used to evaluate transition 
amplitude between any two kinematical states $|x\rangle$ and $|x'\rangle$:
\begin{equation}
W(x,x')=\langle x'|\hat{P}|x\rangle.
\end{equation}
The state $|x\rangle$ might correspond to the initial and $|x'\rangle$ to the final
boundary spin network states (confront with Fig. \ref{SpinFoam}). While the notion 
of the boundary initial and final hypersurfaces is well defined in the case with 
preferred time foliation, the general relativistic case deserves generalization of 
the transition amplitude to the form being independent of the background time 
variable. This leads to the concept of \emph{boundary formulation}  \cite{Oeckl:2003vu} 
of transition amplitudes in which the transition amplitude is a function of boundary 
state only. Taking the particular boundary physical spin network state $|\Psi \rangle$
the transition amplitude can be, therefore, written as     
\begin{equation}
W(\Psi) = \langle W| \Psi \rangle, \label{BoundaryAmp}
\end{equation}
where the state $|\Psi \rangle$ corresponds to representation in which the amplitude 
is evaluated.  

\begin{figure}[ht!]
\centering
\includegraphics[width=8cm,angle=0]{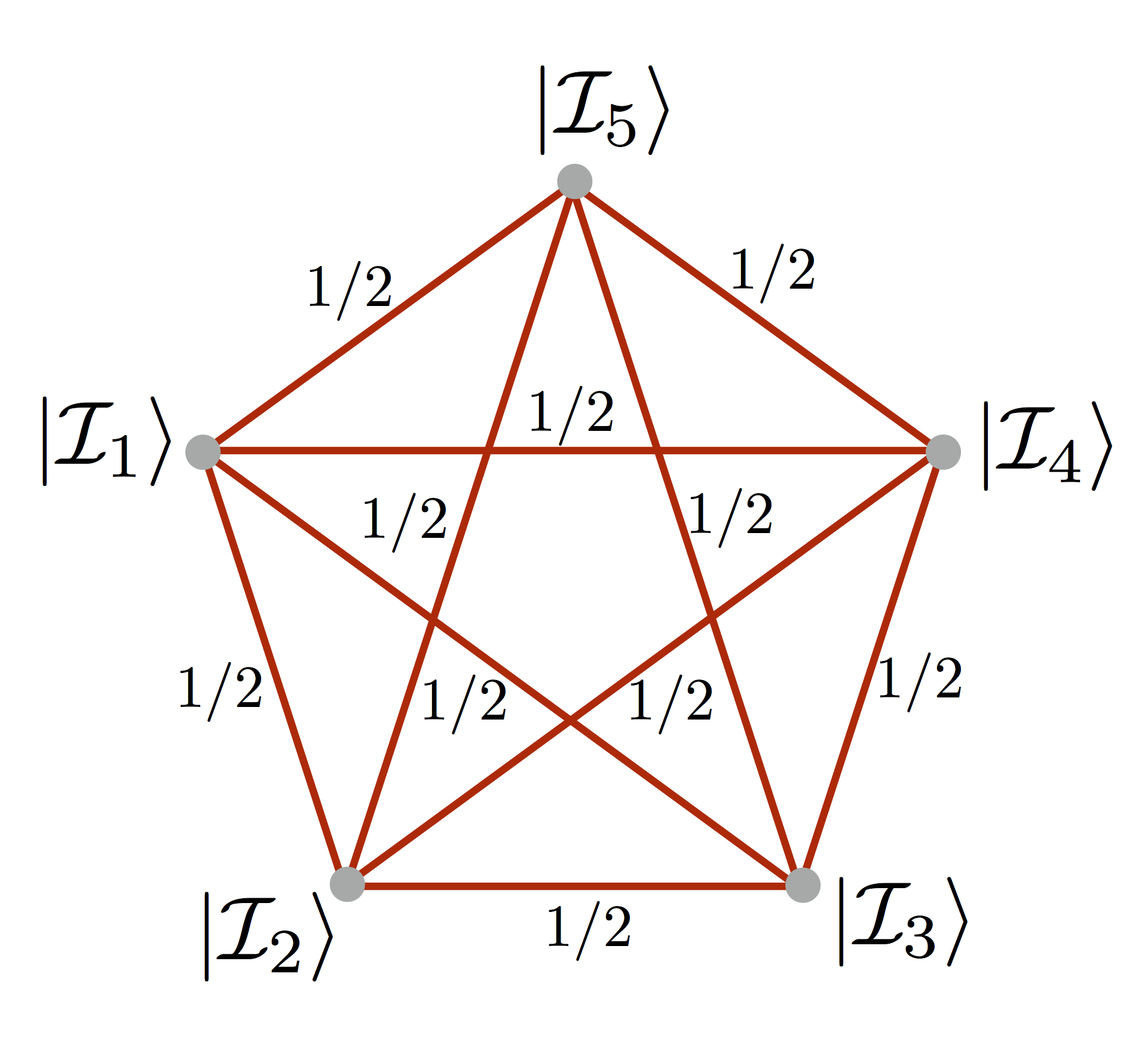}
\caption{Pentagram spin network corresponding to boundary of a single vertex 
of spin foam. The boundary geometry has topology of three-sphere.}
\label{Pentagram}
\end{figure}

The object of our interest in this article, namely the vertex amplitude is the amplitude 
(\ref{BoundaryAmp}) of boundary enclosing a single vertex. As we have already explained
in Introduction, the spin network enclosing the single vertex has pentagram structure 
and can be written as:
\begin{equation}
\langle W|\Psi \rangle = A(\mathcal{I}_1,\mathcal{I}_2,\mathcal{I}_3,\mathcal{I}_4,\mathcal{I}_5).
\label{vertexamplitudeEq}
\end{equation}
The associated spin network is shown in Fig. \ref{Pentagram}.  

The pentagram spin network state is a tensor product of the five intertwiner qubits: 
\begin{equation}
|\Psi \rangle = \bigotimes_{n=1}^5 |\mathcal{I}_{n} \rangle. 
\end{equation}
Since in the vertex amplitude (\ref{BoundaryAmp}) physical states have to be 
considered the intertwiner qubits  $|\mathcal{I}_{n} \rangle$ have to be selected 
such that the state is annihilated by the scalar constraint:  $\hat{C}|\Psi \rangle = 0$. 
Due to the difficulty of the issue for general form of the scalar constraint operator, we 
de not address the problem of selecting $|\Psi \rangle$ states here. As we have 
mentioned, for either symmetry reduced or simplified scalar constraints the physical 
states can be identified with the use of existing methods.  

Another issue is the choice of the state $|W \rangle$. Usually the representation of 
holonomies associated with the links of the spin networks are considered. Here, 
following Ref. \cite{Li:2017gvt} we will evaluate the boundary spin network state in 
the state: 
\begin{equation}
|W \rangle = \bigotimes_{l=1}^{10} |\mathcal{E}_l \rangle,  \label{Wstate}
\end{equation}
where 
\begin{equation}
|\mathcal{E}_l \rangle  = \frac{1}{\sqrt{2}}\left(|01\rangle-|10\rangle \right)  \label{BellState}
\end{equation}
are Bell states associated with the links. Such choice is interesting since the 
Bell states introduce entanglement between faces of the adjacent tetrahedra. 
Such a way of ``gluing" tetrahedra by quantum entanglement has been recently 
studied in Ref. \cite{Baytas:2018wjd}, where it has been shown that 
the $|W \rangle $ state is a superposition of spin network sates (with $\{ 15 j \}$ 
symbols as Clebsch-Gordan coefficients). Since the spin network state $|\Psi \rangle$
is disentangled one can also interpret $\langle W| \Psi \rangle$ as an amplitude 
of transition between disentangled and strongly entangled piece of quantum 
geometry. Going further, possibly the quantum entanglement is the key ingredient 
which merge the chunks of space associated the nodes of spin networks into a 
geometric structure. This reasoning is consistent with the recent advances in the 
domain of entanglement/gravity duality, example of which is provided by the AdS/CFT 
correspondence \cite{Maldacena:1997re}, ER=EPR conjecture \cite{Susskind:2017ney} 
and considerations of holographic entanglement entropy 
\cite{Swingle:2009bg,Ryu:2006bv,Rangamani:2016dms}. Interestingly, it has been recently 
argued that indeed the spin networks may represent structure of quantum entanglement 
\cite{Han:2016xmb}, indicating for relation between spin networks and tensor networks 
\cite{tensornetworks}. This is actually not such surprising since the the holonomies 
associated with links of the spin-networks can be perceived as ``mediators'' of entanglement. 

The holonomies are parallel transport maps between two vector (Hilbert) spaces at 
the ends of a curve $e(\lambda)$, where the affine parameter $\lambda\in[0,1]$. Let 
us denote the initial point as $a=e(0)$ and the final one as $b=e(1)$. Then, holonomy
$h_e$ is a map between two Hilbert spaces $\mathcal{H}^{(a)}$ and  $\mathcal{H}^{(b)}$ 
associated with two (in general) different spatial locations:
\begin{equation}
h_e :  \mathcal{H}^{(a)} \rightarrow \mathcal{H}^{(b)}. 
\end{equation}
In the case considered in this article, the Hilbert spaces are related with the elementary 
qubits $\mathcal{H}_{1/2}$ ``living" at the ends of the links of the spin-networks (keep 
in mind that these are not the intertwiner qubits but the elementary qubits our of which 
the intertwiner qubits are built). As an example of the holonomy of the Ashtekar connection ${\bf A}$ 
considered in LQG (i.e. $h_e :=\mathcal{P} \exp {\int_e {\bf A}}$, see e.g. Ref. \cite{Ashtekar:2004eh}) 
let us consider 
\begin{equation}
h_x(\alpha) := e^{i \sigma_x \alpha} = \mathbb{I} \cos(\alpha)+i \sigma_x \sin(\alpha), \label{holo}
\end{equation} 
where $\alpha$ is an angle variable and $\sigma_x$ is the Pauli matrix. The holonomies as 
the one given by Eq. \ref{holo} are associated with homogeneous models and are consider 
in Loop Quantum Cosmology \cite{Bojowald:2008zzb} and Spinfoam Cosmology 
\cite{Bianchi:2010zs,Vidotto:2010kw}. The special case is when $\alpha=\pi/2$ for which 
$h_x(\pi/2)= i \sigma_x$, which written as an operator 
\begin{equation}
\hat{h}_x(\pi/2)= i \hat{X}, \label{holoop}
\end{equation}
where $\hat{X}$ is the bit-flip operator introduced earlier. Therefore, having e.g. the elementary 
qubit $| 0 \rangle \in \mathcal{H}^{(a)}_{1/2}$ at point $a$, the operator (\ref{holoop}) maps this 
state into $\hat{h}_x(\pi/2)| 0 \rangle =i \hat{X}| 0 \rangle = i |1\rangle \in \mathcal{H}^{(b)}_{1/2}$
at the point $b$ \footnote{Performing an inverse mapping $\hat{h}^{\dagger}_x(\pi/2)$ we can 
map the state $i |1\rangle$ back to $\hat{h}^{\dagger}_x(\pi/2)i |1\rangle =-i \hat{X}i |1\rangle = | 0 \rangle$.}. 
This naturally introduces relation between the quantum states at distant points $a$ and $b$, which 
possibly can be associated with entanglement. In order to illustrate the ``entanglement'' let us consider 
a superposition $|\Psi_a \rangle =\frac{1}{\sqrt{2}}(|0 \rangle+|1\rangle) \in \mathcal{H}^{(a)}_{1/2}$ 
which is mapped into $|\Psi_b \rangle =\frac{i}{\sqrt{2}}(|0 \rangle+|1\rangle) \in \mathcal{H}^{(b)}_{1/2}$. 
If the two states at $a$ and $b$ would be disentangled then performing measurement on the quantum 
state at $a$ would not influence the quantum state at $b$. However, once the measurement is 
performed at the state $|\Psi_a \rangle$ reducing the state to for instance $|0\rangle$, the state at 
$b$ has to be consequently reduced to $i|1\rangle$. The same works in the reverse direction. 
Worth mentioning is that the exemplary correlation via holonomies is consistent with the entanglement 
resulting from the Bell state $|\mathcal{E}_l \rangle$ (\ref{BellState}), which we associated with the links. 
This gives further support to to the choice of the representation state $| W \rangle$ given by Eq. 
\ref{Wstate}. However, the issue of relation between holonomies and entanglement requires further 
more detailed studies, also in the spirit of the recent proposal of \emph{Entanglement holonomies} 
\cite{Czech:2018kvg}.  In particular, it has to be confirmed that the correlation introduced via holonomies 
is the true quantum entanglement violating Bell inequalities. 

\section{A quantum algorithm} \label{Algorithm}

Having the vertex amplitude (\ref{vertexamplitudeEq}) defined we may proceed to the 
task of determining $|\langle W |\Psi \rangle|^2$ with the use of quantum computers.
Here, we will show how to obtain modulus the amplitude modulus square (the probability) 
while extraction of the phase factor will be a subject of our further investigations.  

Let us begin with preparation of a suitable quantum register. Because each of the intertwiner 
qubits is a superposition of four elementary qubits, evaluation of the spin network with $N$ 
nodes requires $4N$ qubits in the quantum register\footnote{This statement is made under 
assumption that no ancilla qubits are required.}. The corresponding Hilbert space is spanned 
by $2^{4N}$ basis states $| {\bf i} \rangle$, where ${\bf i} \in \left\{ 0, \dots, 2^{4N}-1 \right\}$. 
The initial state for the quantum algorithm is:   
\begin{equation}
| {\bf 0} \rangle = \underbrace{| 0 \rangle \otimes \dots \otimes| 0 \rangle}_{4N}.  \label{InitialState}
\end{equation}

Now, we have to find unitary operators $\hat{U}_{\Psi}$ and $\hat{U}_{W}$ defined such that 
\begin{align}
|\Psi \rangle &= \hat{U}_{\Psi}| {\bf 0} \rangle, \\
|W \rangle &= \hat{U}_{W}| {\bf 0} \rangle,
\end{align}
where $| {\bf 0} \rangle$ is given by Eq. \ref{InitialState}. Utilizing the operators $\hat{U}_{\Psi}$ 
and $\hat{U}_{W}$ we introduce an operator $\hat{U}:=\hat{U}_{W}^{\dagger}\hat{U}_{\Psi}$. 
Action of this operator on the initial state (\ref{InitialState}) can be expressed as a superposition 
of the basis states with some amplitudes $a_i \in \mathbb{C}$:
\begin{equation}
\hat{U}| {\bf 0} \rangle = \sum_{i=0}^{2^{4N}-1} a_i | {\bf i} \rangle. 
\end{equation}
It is now easy to show that the $a_0$ coefficient in this superposition is the transition amplitude 
we are looking for. Namely:
\begin{equation}
a_0 = \langle {\bf 0}|\hat{U}| {\bf 0} \rangle = 
\langle {\bf 0}|\hat{U}_{W}^{\dagger}\hat{U}_{\Psi}| {\bf 0} \rangle 
= \langle W|\Psi \rangle. 
\end{equation}
By performing measurements on the final state we find the probabilities $P(i)=|a_i|^2$. The 
first of these probabilities is the modulus square of the vertex amplitude. 

Before we will proceed to the discussion of the pentagram spin network associated with the 
vertex amplitude let us first demonstrate the algorithm on two simpler examples of spin 
networks with one and two nodes.  

\subsection{Example 1 - single tetrahedron}

\begin{figure}[ht!]
\centering
\includegraphics[width=6cm,angle=0]{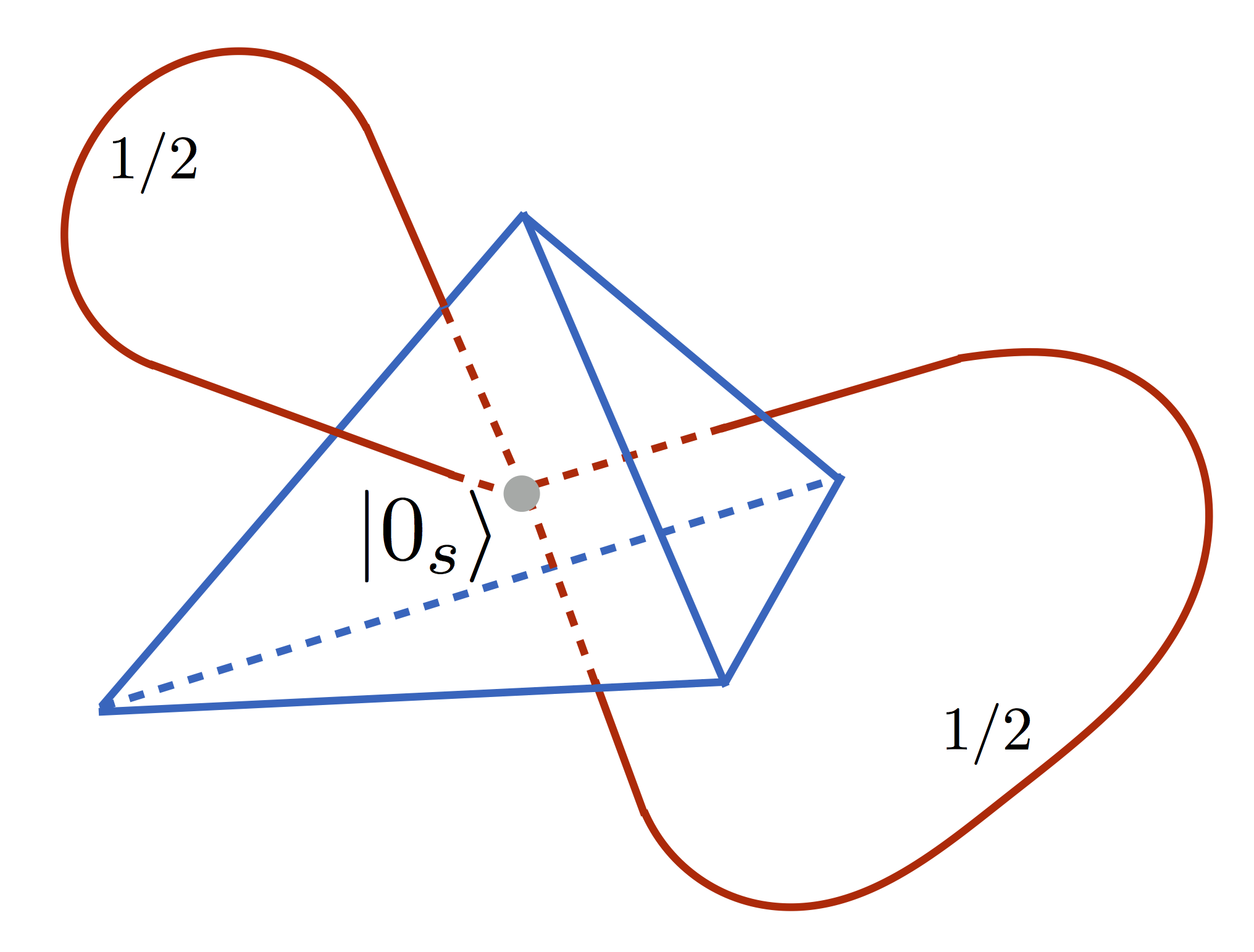}
\caption{A single-node spin network associated with identification of the pairs of face of a tetrahedron.}
\label{SingleNode}
\end{figure}

As a first example let us consider the case of a single-node spin network presented in 
Fig. \ref{SingleNode}.
Here, the intertwiner qubit is in the $| \Psi \rangle = | 0_s \rangle$ state composed out of the four 
elementary qubits according to Eq. \ref{0sstate}. The representation state $| W \rangle$ 
is a tensor product of two Bell states  (\ref{BellState}). There are basically two different choices
of pairing faces of the tetrahedron. The first choice is according to the pairing of qubits entering 
to the two-qubit singlets $| S \rangle $ out of which the $ | 0_s \rangle$ state is built. The second
choice is by linking qubits contributing to the two different singlets. The first choice is trivial 
since in that case $\hat{U}_{W} = \hat{U}_{\Psi}$ and in consequence the amplitude 
$\langle W|\Psi \rangle = \langle {\bf 0}|\hat{U}_{W}^{\dagger}\hat{U}_{\Psi}| {\bf 0} \rangle = 
\langle {\bf 0}| {\bf 0} \rangle =1$. Therefore, we will consider the second case for which the 
quantum circuit associated with the $\hat{U} = \hat{U}_{W}^{\dagger}\hat{U}_{\Psi}$ operator 
is presented in Fig. \ref{SingeNodeCirc}.
 
\begin{figure}[ht!]
\centering
\includegraphics[width=16cm,angle=0]{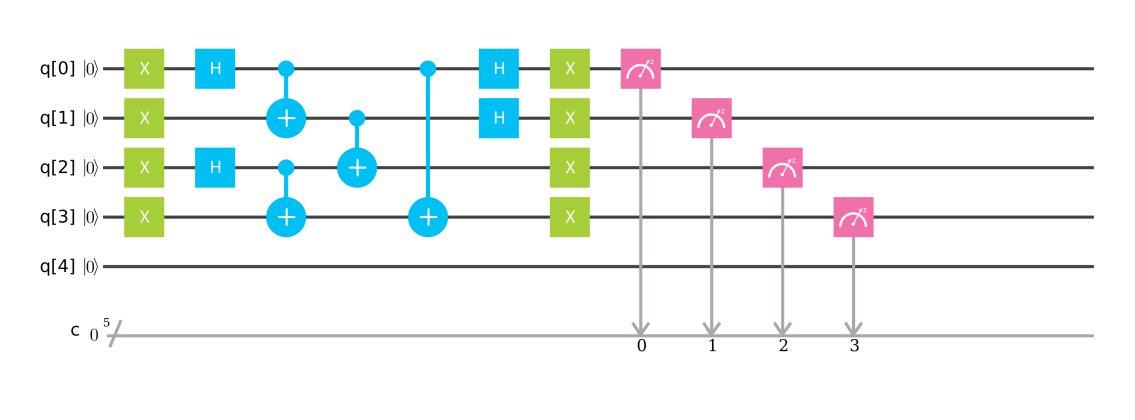}
\caption{Quantum circuit used to evaluate $|\langle W|\Psi \rangle|^2$ the boundary transition
amplitude of the spin network presented in Fig. \ref{SingleNode}.}
\label{SingeNodeCirc}
\end{figure} 

The simulations were performed on both the IBM simulator and the QX simulator, 
with 1024 shots in each computational round. The rounds have been repeated 10 
times. The results obtained are collected in Table \ref{Table2}.
 
\begin{table}[h!]
\begin{tabular}{|c|c|c|c|c|}
\hline
No. &\  $P_0$ (QX)   \   &\ Hits of $|{\bf{0}}\rangle$ (QX) &  \ $P_0$ (IBM)   \   
&\ Hits of $|{\bf{0}}\rangle$ (IBM)  \\ 
\hline
1   & 0.255859375 &  262 &  0.263671875 & 270  \\
2   & 0.248046875 &  254  & 0.2529296875  & 259 \\
3   & 0.267578125 &  274  & 0.2578125  & 264 \\
4   & 0.2568359375 &  263 & 0.27734375   & 284  \\
5   & 0.2568359375 &  263 & 0.232421875  & 238  \\
6   & 0.25 &  256                &  0.263671875 & 270   \\
7   & 0.2578125 & 264       &  0.25 & 256   \\
8   & 0.23828125 &  244   &  0.244140625 & 250   \\
9   & 0.240234375 &  246  & 0.2607421875  & 267  \\
10 & 0.25 &  256                 & 0.2763671875  & 283   \\ 
\hline
\end{tabular}
\caption{Results of measurements of $P(0)=|a_0|^2$ for the quantum circuit 
presented in Fig. \label{SingeNodeCirc}, using both the IBM simulator and the 
QX simulator. Each measurement corresponds to the number of shots equal 
1024.}
\label{Table2}
\end{table}

Averaging over the ten rounds the following values of modulus square of the 
amplitudes are obtained: 
\begin{equation}
|\langle W|\Psi \rangle|^2 =|a_0|^2 = \left\{ \begin{tabular}{cc}  
0.252 $\pm$ 0.009 & \text{for QX simulator} \\  
0.256 $\pm$ 0.014 & \text{for IBM simulator}  \end{tabular} \right. .
\end{equation}
The results are consistent with the theoretically expected value $|a_0|^2 = 0.25$. 
Finally, worth mentioning is that the algorithm cannot directly be executed using the 
IBM Q 5-qubit quantum chip due to the topological constraints of the structure of 
coupling between qubits. Additional ancilla qubits would have to be involved for this 
purpose.

\subsection{Example 2 - two tetrahedra}

\begin{figure}[ht!]
\centering
\includegraphics[width=8cm,angle=0]{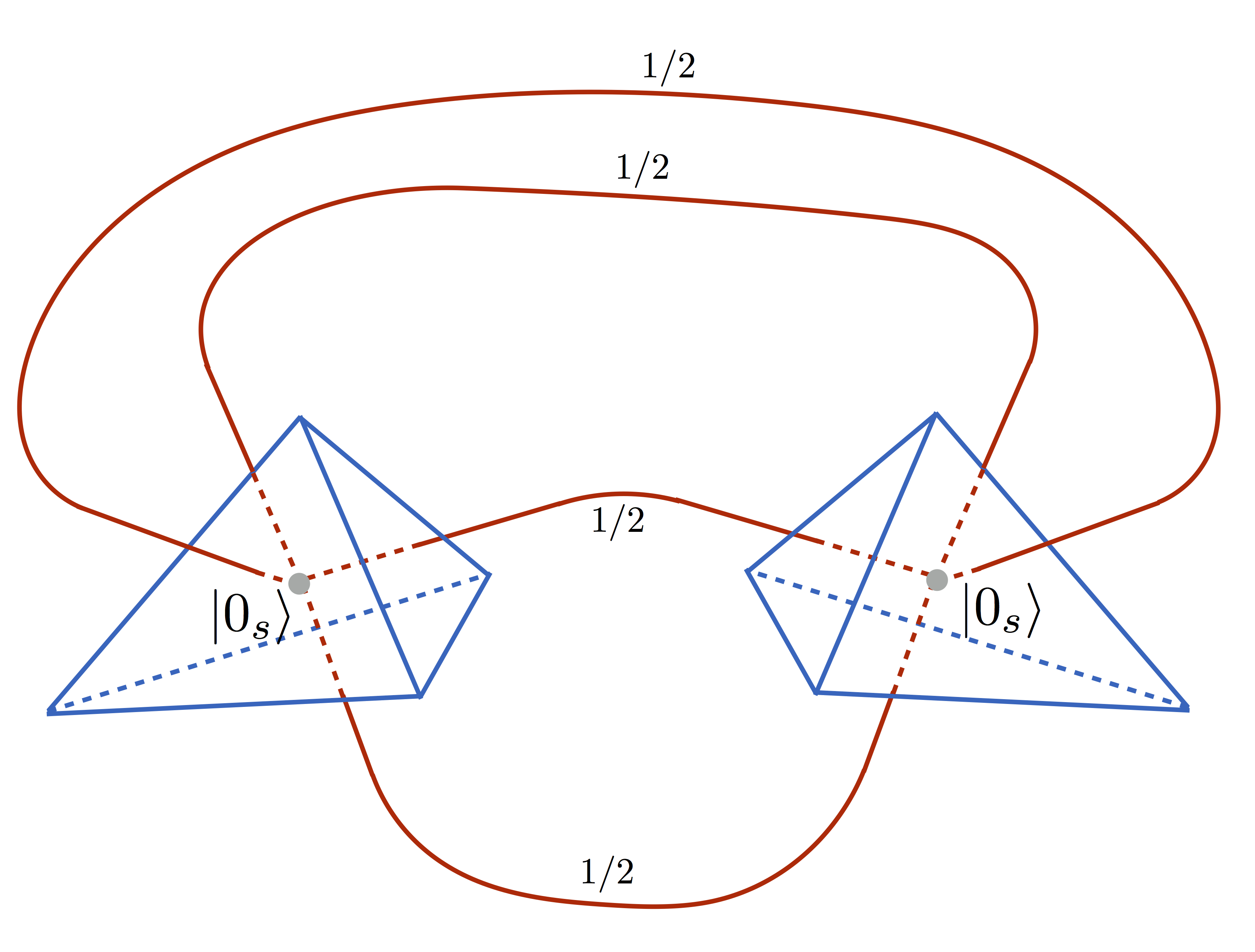}
\caption{A two-node spin network associated with two tetrahedra.}
\label{TwoNodes}
\end{figure}

The second example concerns a bit more complex situation with two-node spin 
network presented in Fig. \ref{TwoNodes}.
Here, the representation state $|W\rangle$ similarly to the previous example is 
associated with the Bell sates (\ref{BellState}) entangling faces of the two tetrahedra
one into anther. The corresponding choice of the quantum circuit used the evaluate 
the boundary amplitude is presented in Fig. \ref{TwoNodesCirc}.
\begin{figure}[ht!]
\centering
\includegraphics[width=16cm,angle=0]{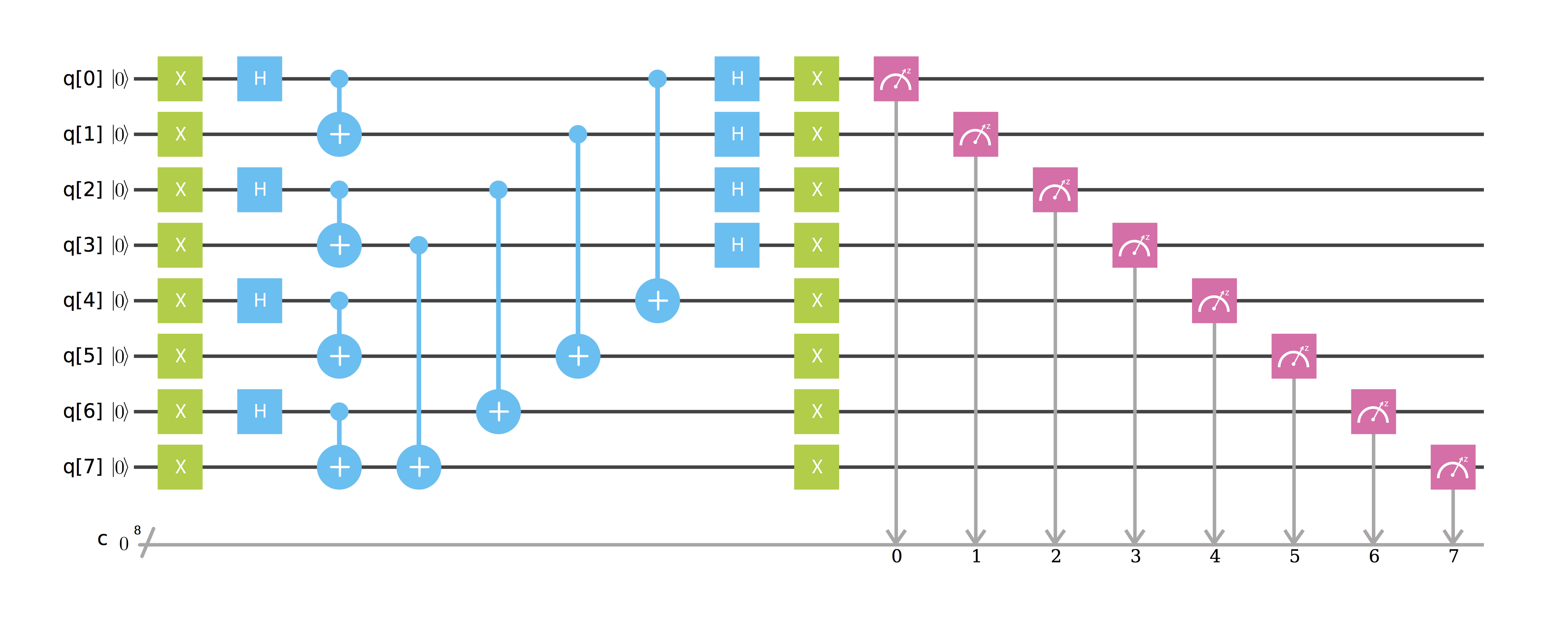}
\caption{Quantum circuit used to evaluate $|\langle W|\Psi \rangle|^2$ the boundary 
transition amplitude of the spin network presented in Fig. \ref{TwoNodes}. The qubits 
$\{0,1,2,3\}$ belong to the one node while the qubits $\{4,5,6,7\}$ belong to the another.
The links are between the pairs of qubits: $\{0,4\}$, $\{1,5\}$, $\{2,6\}$ and $\{3,7\}$.}
\label{TwoNodesCirc}
\end{figure}

The simulations were performed on both the IBM simulator and the QX simulator, 
with 1024 shots in each round. As in the previous example, the computational rounds 
have been repeated 10 times.  The results obtained are collected in Table \ref{Table3}.

\begin{table}[h!]
\begin{tabular}{|c|c|c|c|c|}
\hline
No. &\  $P_0$  (QX) \   &\ Hits of $|{\bf{0}}\rangle $ \ (QX)  &\  $P_0$  (IBM) \   
&\ Hits of $|{\bf{0}}\rangle $ (IBM) \   \\ \hline
1   & 0.0595703125 &  61 & 0.0556640625 & 57 \\
2   & 0.0595703125 &  61 & 0.06640625 & 68 \\
3   & 0.06640625    &  68  & 0.060546875 & 62  \\
4   & 0.0615234375 &  63  & 0.064453125 & 66  \\
5   & 0.080078125  &  82  & 0.0634765625 &  65 \\
6   & 0.0498046875 &  51 & 0.0595703125 &  61  \\
7   & 0.052734375  &  54 &  0.0615234375 &  63  \\
8   & 0.072265625  & 74  & 0.0537109375 &  55 \\
9   & 0.0673828125 &  69 & 0.052734375 &  54  \\
10 & 0.0634765625 &  65 & 0.0654296875 & 67   \\ 
\hline
\end{tabular}
\caption{Results of measurements of $P(0)=|a_0|^2$ for the quantum circuit 
presented in Fig. \label{TwoNodesCirc} using both the IBM simulator and the 
QX simulator. Each measurement corresponds to the number of shots equal 
1024.}
\label{Table3}
\end{table}

Performing averaging over the computational rounds the following values 
of $|\langle W|\Psi \rangle|^2$ are obtained: 
\begin{equation}
|\langle W|\Psi \rangle|^2 =|a_0|^2 = \left\{ \begin{tabular}{cc}  
0.063 $\pm$ 0.009 & \text{for QX simulator} \\  
0.060 $\pm$ 0.005 & \text{for IBM simulator}  \end{tabular} \right. .
\end{equation}
The results are in agreement with the theoretically expected value $|\langle W|\Psi \rangle|^2 
=|a_0|^2 =0.625$ (obtained using Quirk \cite{Quirk}).

\section{Evaluation of vertex amplitude} 

We are now ready to address the task of determining the vertex amplitude 
(\ref{vertexamplitudeEq}) associated with the boundary spin network state: 
\begin{equation}
|\Psi \rangle = |0_s \rangle \otimes |0_s \rangle\otimes |0_s \rangle\otimes 
|0_s \rangle\otimes |0_s \rangle. 
\end{equation}
The other possible choices of the spin network state will discussed in our 
further work \cite{Mielczarek2019}. 

\begin{figure}[ht!]
\centering
\includegraphics[width=16cm,angle=0]{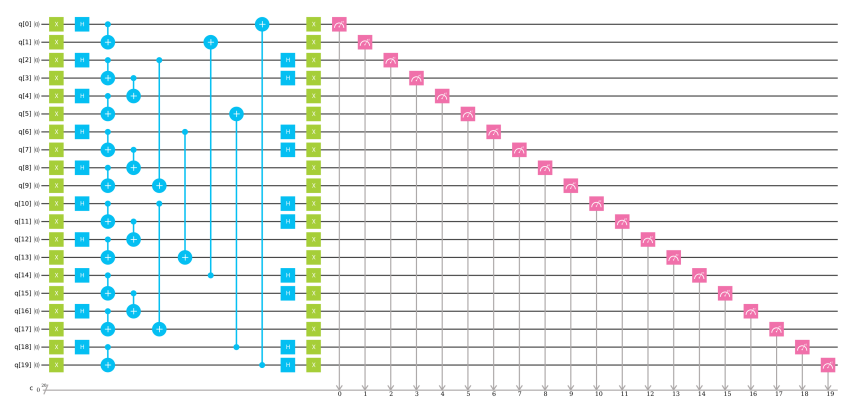}
\caption{Quantum circuit used to determine $|A(0_s,0_s,0_s,0_s,0_s)|^2$. Nodes of the spin 
network correspond to the following sets of qubits: $\{0,1,2,3 \}$, $\{4,5,6,7 \}$, $\{8,9,10,11 \}$, 
$\{12,13,14,15 \}$, $\{16,17,18,19 \}$. The links are between the pairs of qubits: 
$\{0,19\}$, $\{1,14\}$, $\{2,9\}$, $\{3,4\}$, $\{5,18\}$, $\{6,13\}$, $\{7,8\}$, $\{10,17\}$, $\{11,12\}$ 
and $\{15,16\}$.}
\label{VertexAmplitudeCirc}
\end{figure}

The $|W\rangle$ is given by Eq. \ref{Wstate}, representing entanglement between faces 
of tetrahedra being connected by the links of the spin network. Due to anti-symmetricity 
of the Bell states (\ref{BellState}) for the $10$ links under consideration we have in 
general $2^{10}=1024$ ways to order the states between the nodes of the spin network. 
Here, in order to not distinguish any of the nodes, the configuration in which every node 
is entangled with two other nodes by the state $|\mathcal{E}_l \rangle  = 
\frac{1}{\sqrt{2}}\left(|01\rangle-|10\rangle \right)$ and another two nodes by the state  
$e^{i\pi}|\mathcal{E}_l \rangle  = \frac{1}{\sqrt{2}}\left(|10\rangle-|01\rangle \right)$ is considered.  
The resulting quantum circuit corresponding to the operator 
$\hat{U} = \hat{U}_{W}^{\dagger}\hat{U}_{\Psi}$, together with the measurements necessary 
to find $|a_0|^2=|\langle W|\Psi \rangle|^2$ is shown in Fig. \ref{VertexAmplitudeCirc}.

The quantum circuit employs 20-qubit quantum register with the initial state: 
\begin{equation}
| {\bf 0} \rangle = \bigotimes_{n=0}^{19} | 0 \rangle. \label{instate}
\end{equation}

The algorithm introduced in Sec. \ref{Algorithm} requires finding amplitude of the  
initial state (\ref{instate}) in the final state. One has to keep in mind that the Hilbert 
space of the 20-qubit system is spanned by over one million basis states: 
$2^{20} = 1048576$. Therefore, selecting amplitude of one of the basis states 
(i.e. $| {\bf 0} \rangle$) is not an easy task. 

The first attempt to determine the value of $|\langle W|\Psi \rangle|^2 =|A(0_s,0_s,0_s,0_s,0_s)|^2$
has been made with us of the IBM simulator of quantum computer. Ten rounds of simulation,  
each of $1024$ shots, have been performed. However, no single event with the $| {\bf 0} \rangle$
state in the final state has been observed. Assuming that the probability is evenly distributed 
between the basis states the probability $1/2^{20} \approx 10^{-6}$ per basis states can be expected. 
With the $10240$ measurements made, this gives roughly $1\%$ chance to observe the state. 

The second attempt to determine value of the vertex amplitude has been made with the use of the 
QX quantum computer simulator. Similarly to the simulations performed on the IBM simulator, ten 
computational rounds, each of $1024$ shots, have been performed. In this case, the events with 
$| {\bf 0} \rangle$ have been observed and are collected in Table \ref{Table4}.

\begin{table}[h!]
\begin{tabular}{|c|c|c|}
\hline
No. &\  $P_0$  \   &\ Hits of $|{\bf{0}}\rangle $ \    \\ \hline
1   &  0.0009765625 &  1    \\
2   &  0.0029296875 &  3    \\
3   &  0                      &  0    \\
4   &  0.0009765625 &  1    \\
5   &  0.001953125   &  2    \\
6   &  0.0009765625 &  1    \\
7   &  0.001953125   &  2    \\
8   &  0.0009765625 &  1    \\
9   &  0.0029296875 &  3    \\
10 &  0.0009765625 &  1    \\ 
\hline
\end{tabular}
\caption{Results of measurements of $P(0)=|a_0|^2$ for the quantum circuit 
presented in Fig. \ref{VertexAmplitudeCirc} using the QX simulator. Each 
measurement corresponds to the number of shots equal 1024.}
\label{Table4}
\end{table}
By averaging the results from Table \ref{Table4}, the following value of the modulus 
square of the vertex amplitude $\langle W|\Psi\rangle$ can be found: 
\begin{equation}
|\langle W|\Psi \rangle|^2 =|A(0_s,0_s,0_s,0_s,0_s)|^2 = P_0 = 0.00147 \pm 0.00095. 
\label{FinalResult}
\end{equation}

The results obtained from the IBM and QX simulators are contradictory. However, the 
QX simulator result (\ref{FinalResult}) is much closer to the theoretically expected value. 
Namely, the spin foam amplitude considered in this section can be determined using 
recoupling theory for $SU(2)$ group. Following the discussion in Ref. \cite{Baytas:2018wjd} 
on can find that the amplitude (\ref{vertexamplitudeEq}) is given by the $\{15j\}$ symbol.
Employing definition of the symbol (see e.g. Eq. 17 in Ref. \cite{Dona:2017dvf}) 
for all the spin labels $j_{ab}=1/2$ and the intertwiners $i_a=0$ (which correspond 
to the $|0_s\rangle$ states), where $a,b \in\{1,2,3,4,5\}$, one can find that $\{15j\}=0.0625$. 
In conserquence $|\langle W|\Psi \rangle|^2 =|\{15j\}|^2 = 0.0625^2 = 0.00390625$.
The difference between this prediction and the result of simulation (\ref{FinalResult})
goes beyond the statistical error and, therefore, one can expect that systematic error 
was involved. Resolution of this issue requires further investigation, especially analysis 
of the sampling methods used in the quantum simulator. The same concerns 
the IBM quantum simulator where discrepancy between the theoretically predicted 
value and the results of measurements is more serious. 

Taking the above into account, a comment on utility of the applied methodology is desirable. 
First of all, we already used the fact that the vertex amplitude discussed in this section 
can easily be evaluated without the need of quantum circuits. The vertex amplitude,  
associated with the pentagram spin network is given by the $\{15j\}$ symbol. Recently, 
significant progress has been made in the development of numerical methods of evaluation 
of spin foam vertex amplitudes \cite{Sarno:2018ses,Dona:2018nev,Dona:2019dkf}.
However, there are still some obstacles, e.g. oscillating nature of the spin foam amplitudes, 
which motivate search for alternative computational methods.     

The aim of our study was to provide the proof of the concept of applicability of quantum circuits 
to determine quantities being of relevance in Loop Quantum Gravity and Spin Foam approaches. 
We have shown that indeed, despite of the current hardware limitations (see Appendix A), 
interesting quantities can already be evaluated on quantum computer simulators. As we 
demonstrated, 2 qubits per link of a spin network are needed for this purpose. Therefore, in 
case of the 5-node pentargam spin network (with 10 links) studied here, 20 qubit register was 
used. Utilizing the available commercial quantum simulators (e.g. $37$ qubit QX multi-node 
simulator SurfSara \cite{QuantInsp}), amplitudes of spin networks with up to 9 4-valent nodes 
can potentially be computed. 

Our plant is to perform such simulations in our further studies. Furthermore, our goal is to extend 
computational capabilities to $40$ qubits using academic and commercial supercomputing resources. 
This will allow to simulate spin networks with 10 nodes. Worth mention is that while scaling the system size 
is straightforward for the method based on quantum circuits, the application of the standard methods 
based on recoupling theory may turn out to be inefficient. This will become even more evident when 
advantageous quantum computers will become available (as expected) in the second half on the 
coming decade (see Appendix A), providing 100 and more fault tolerant qubits. Until that time there
is still a lot of potential for improving simulator-based computations and development of methods 
which will ultimately be applied on real quantum processors.
 
Furthermore, the quantum simulations will not only allow to study amplitudes but also other 
relevant quantities. In particular, analysis of quantum fluctuations and quantum entanglement 
between subsystems will be possible to investigate. One of the open problems which will 
become possible to study by extending the methodology introduced here is the entanglement 
entropy between subsystems of spin networks and the issue of area law for entropy of entanglement. 
 
\section{Summary}

The purpose of this article was to explore the possibility of computing vertex amplitudes 
in the spin foam models of quantum gravity with the use of quantum algorithms. The 
notion of \emph{intertwiner qubit} being crucial to implement the vertex amplitudes on 
quantum computers has been pedagogically introduced. It has been shown how one of the
two basis states of the intertwiner qubit can be implemented with the use available IBM 5-qubit 
quantum computer. To the best of our knowledge it was the first time ever a 
quantum gravitational quantity has been simulated on superconducting quantum chip.

Thereafter, a quantum algorithm allowing to determine modulus square of 
spin foam vertex amplitude ($|\langle W|\Psi \rangle|^2$) has been introduced. Utility of the 
algorithm has been demonstrated on examples of single-node and two-node spin networks. 
For the two cases, probabilities of the associated boundary states have been determined with 
the use of IBM and QX quantum computer simulators. Finally, the algorithm has been applied 
to the case of pentagram spin network, representing boundary of the spin foam vertex. 
Value of the modulus square of the amplitude in a certain quantum state has been 
measured with the use of 20-qubit register of the IBM and QX quantum simulators. While the 
QX results were close to the analytically predicted value, the outcomes of the IBM simulator 
failed to reproduce theoretical predictions.

The presented results are the first step towards simulating spin foam models (associated 
with Loop Quantum Gravity and Group Field Theories) with the use of universal quantum 
computers. In particular, the vertex amplitudes can be applied as elementary building blocks 
in construction of more complex transition amplitudes. The aim of the developed direction 
is to achieve possibility of studying collective behavior of the Planck scale systems composed 
of huge number of elementary constituents (``atoms of space/spacetime"). Exploration of 
the many-body Planck scale quantum systems \cite{Oriti:2017twl} may allow to extract 
continuous and semi-classical limits from the dynamics of the ``fundamental" degrees of 
freedom. This is crucial from the perspective of making contact between Planck scale physics 
and empirical sciences.  

Worth stressing is that the results presented in this article are rather preliminary and only 
set up the stage for further more detailed studies. In particular, the following points have 
to be addressed:
\begin{itemize}
\item Introduction of a quantum circuit for the general intertwiner qubit $|\mathcal{I}\rangle$ 
(Eq. \ref{IntertwinerState}).
\item Determination of the phase of vertex amplitude with the use of quantum algorithms 
(e.g. Quantum Phase Estimation Algorithm \cite{Cleve:1997dh}).
\item Investigation of different types of the state $|W \rangle$. 
\item Analysis of spin networks with up to $10$ nodes on quantum computers simulators.   
\item A possibility of solving quantum constraints with the use of quantum circuits. 
\item Application to Spinfoam cosmology \cite{Bianchi:2010zs,Vidotto:2010kw}.
\item Investigation of the architectures of forthcoming quantum processors (with the 
$N>100$ number of qubits) in terms of application to spin foam transition amplitudes.  
\end{itemize}
Some of the tasks will be subject of a sequel to this article \cite{Mielczarek2019}.   

\section*{Acknowledgements}

JM is supported by the Sonata Bis Grant DEC-2017/26/E/ST2/00763 of the National Science 
Centre Poland and the Mobilno\'s\'c Plus Grant 1641/MON/V/2017/0 of the Polish Ministry of 
Science and Higher Education. 

\section*{Appendix A - Quantum computing technologies}

The domain of quantum computing is currently experiencing an unprecedented speedup. 
The recent progress is mainly due to advances in development of the superconducting qubits 
\cite{You}. In particular, utilizing the superconducting circuits, the IBM company has developed 
5 and 20 qubit (noisy) quantum computers, which are accessible in cloud \cite{IBM}. Furthermore, 
a prototype of 50 qubit quantum computer by this company has been built and is currently in the 
phase of tests. The company has recently also unveiled its first commercial 20-qubit quantum 
computer IBM Q System One. This is intended to be the first ever commercial universal quantum 
computer, and the second commercially available quantum computer after the adiabatic quantum 
computer (quantum annealer \cite{Kadowaki}) provided by D-Wave Systems \cite{DWave}. The 
latests D-Wave 2000Q annealer uses a quantum chip with 2048 superconducting qubits, connected 
in the form of the so-called \emph{chimera} graph. Another important player in the quantum race 
is Intel, which recently developed its 49-qubit superconducting quantum chip named Tangle 
Lake \cite{TangleLake}. However, outside of the superconducting qubits, the company in collaboration 
with QuTech \cite{QuTech} advanced centre for Quantum Computing and Quantum Internet 
is also developing an approach to quantum computing based on electron's spin based qubits, 
stored in quantum dots. Further advances in the area of superconducting quantum circuits come 
from Google \cite{Google} and Rigetti Computing \cite{Rigetti}. The first company has recently 
announced their 72-qubit quantum chip, while the second one is currently developing its 128-qubit 
universal quantum chip. On the other hand, the world's leading software company - Microsoft has 
focused its approach to quantum computing on topological qubits through Majorana fermions 
\cite{Microsoft}. The alternative to superconducting qubits is also developed by IonQ Inc. 
startup, which is developing a trapped ion quantum computer based on ytterbium atoms  \cite{Ionq}. 
The most recent quantum computer by this company allows to operate on 79 qubits, which is the 
current world record. The above are only the most sound examples of the advancement which has 
been made in the recent years in the area of hardware dedicated to quantum computing. 
There are still many challenges to be addressed, including reduction of the gate errors and increase 
of fidelity of the quantum states. However, even in pessimistic scenarios, the current momentum 
of the quantum computing technologies will undoubtedly lead to emergence of reliable and 
advantageous quantum machines (which cannot be emulated on classical supercomputers) in 
the coming decade. There are no fundamental physical reasons identified, which could stop the 
progress. However, the rate of the progress will depend on whether commercial applications
of quantum computing technologies will emerge in the coming 5 years, stimulating further funding 
of research and development. See e.g. Ref. \cite{Report} for more detailed discussion of this issue. 

Major players in the field, with large financial resources, such as IBM or governments
may sustain the progress independently on short-term returns (which is not the case for start-ups). 
This may allow for a stable long term progress. In particular, the IBM has recently announced 
a possibility of doubling a measure called \emph{Quantum Volume} \cite{QuantVol} every 
year \cite{IBMQuantVol}. The Quantum Volume $V_Q$ is basically a maximal size of a certain 
random circuit, with equal width and depth, which can be successfully implemented on a given 
quantum computer. The current (2019) IBM's value of $V_Q$ is 16 and corresponds to the IBM 
Q System One quantum computer mentioned above. This means that any quantum algorithm 
employing 4 qubits and 4 layers (time steps) of quantum circuit can be successfully implemented on the 
computer. If the trend will follow the hypothesized geometric trajectory (a sort of a new Moore's 
law \cite{Moore} for integrated circuits), then one could expect the quantum volume $V_Q$ to be 
of the order of $10^3$ in 2025 and $10^5$ in 2030. This means that in 2025 algorithms 
employing roughly $\log_2 10^3 \approx 10$ qubits and the same number of time steps will 
be possible to execute. This number will increase to rapproximately 16 qubits till the end of the 
coming decade.  While this may not sound very optimistic, the prediction is very conservative 
and does not rule out that much bigger (non-random) circuits (especially well-fitted to the 
hardware) will be possible to execute at the same time.    

\section*{Appendix B - Basics of quantum computing}

The aim of the appendix is to provide a basic introduction of the concepts in quantum 
computing used in this article. This appendix will allow quantum gravity researchers 
who are not familiar with quantum computing, to grasp the relevant concepts.   

The quantum computing is basically processing of quantum information. While the 
elementary portion of classical information is a \emph{bit} $\{ 0,1 \}$, its quantum 
counterpart is what we call a \emph{qubit}.  A single qubit is a state $| \Psi \rangle$  
in two-dimensional Hilbert space, which we denote as $\mathcal{H} = \text{span}
\{ |0 \rangle, | 1 \rangle \}$. The space is spanned by two orthonormal basis 
states $|0\rangle $ and $|1\rangle $, so that $\langle 1|0\rangle =0 $ and $\langle 0|0\rangle 
= 1 = \langle 1|1\rangle $. A general qubit is a superposition of the two basis 
states:
\begin{equation}
| \Psi \rangle = \alpha|0\rangle +\beta |1\rangle,
\end{equation}
where, $\alpha, \beta \in \mathbb{C}$ (complex numbers), and the normalization condition
$\langle \Psi | \Psi \rangle = 1$ implies that $ |\alpha|^2+|\beta|^2=1$.

There are different unitary quantum operations which may be performed on the quantum 
state $| \Psi \rangle$. The elementary quantum operations are called \emph{gates}, in 
analogy to electric circuits implementing Boolean logic. For instance, the so-called bit-flip 
operator $\hat{X}$ which transforms  $|0\rangle$ into $|1\rangle$ and  $|1\rangle$ into 
$|0\rangle$  ($\hat{X}|0\rangle =|1\rangle$ and $ \hat{X}|1\rangle =|0\rangle$) can be 
introduced. The $\hat{X}$ operator introduces the NOT operation on a single qubit, 
and has representation in the form of the Pauli $x$ matrix. Similarly, one can introduce 
$\hat{Y}$ and $\hat{Z}$ operators corresponding to the other two Pauli matrices. 
The \emph{computational basis} $\{ |0 \rangle, | 1 \rangle \}$ is usually introduced such that 
the basis states are eigenvectors of the $\hat{Z}$ operator:   $\hat{Z}  |0 \rangle = |0 \rangle$
and $\hat{Z}  |1 \rangle = - |1 \rangle$. 
 
Another important operator (which does not have its classical counterpart) is the Hadamard 
operator $\hat{H}$ which is defined by the following action on the qubit basis states:
\begin{equation}
\hat{H}|0\rangle =\frac{1}{\sqrt{2}}\left(|0\rangle  +|1\rangle\right) \ \text{and} \ 
\hat{H}|1\rangle =\frac{1}{\sqrt{2}}\left(|0\rangle  -|1\rangle\right).
\end{equation}

The above are examples of operators acting on a single qubit. However, quantum information
processing usually concerns a multiple qubit system called \emph{quantum register}. 
The quantum state of the register of $N$ qubits belongs to a tensor product of $N$
copies of single qubit Hilbert spaces: $\bigotimes_{i=1}^N\mathcal{H}_i$. The dimension 
of the product Hilbert space is $\text{dim}\left(\bigotimes_{i=1}^N\mathcal{H}_i  \right)=2^N$. 
This exponential dependence of the dimensionality on $N$ is the main obstacle behind simulating 
quantum systems on classical computers. 

A \emph{quantum algorithm} is a unitary operator $\hat{U}$ acting on the initial state 
of the quantum registes $|{\bf 0}\rangle: =  \otimes_{i=1}^N|0\rangle  \in \bigotimes_{i=1}^N\mathcal{H}_i$,
together with a sequence of measurements.  The outcome of the quantum algorithm is obtained 
by performing measurements on the final sate: $ \hat{U}|{\bf 0}\rangle$. Because of the probabilistic 
nature of quantum mechanics, the procedure has to be performed repeatedly in order to reconstruct 
the final state. In general, full reconstruction of the final state $\hat{U}|{\bf 0}\rangle$ requires 
the so-called \emph{quantum tomography} to be applied. In the procedure, states 
of the qubits are measured in different bases (not only in the computational basis).
The quantum state tomography, which reconstructs the density matrix $\hat{\rho}=\hat{U}|{\bf 0}\rangle\langle{\bf 0} 
|\hat{U}^{\dagger}$, is however not always required. In most of the considered 
quantum algorithms, only probabilities (not complex amplitudes) of the basis states are 
necessary to measure, which is much simpler and faster than the quantum state tomography.

As already mentioned, the unitary operator $ \hat{U}$ can be decomposed into elementary 
operations called quantum gates. The already introduced $\hat{X}$ and $\hat{H}$ operators 
are examples of single-qubit gates. However, the gates may also act on two or more qubits. 
An example of 2-qubit gate relevant for the purpose of this article is the so-called controlled-NOT 
(CNOT) gate, which we denote as $\hat{C}$.  The operator is acting on 2-qubit state 
$|ab\rangle \equiv |a\rangle \otimes|b\rangle$, where $|a\rangle$ and $ |b \rangle$ 
are single quibit states. Action of the CNOT operator on the basis states can be expressed 
as follows: $\hat{C}(|a\rangle \otimes|b\rangle) =|a\rangle \otimes|a\oplus b\rangle$,
where $a,b \in \{0,1\}$.  The $ \oplus$ is the XOR (exclusive or) logical operation 
(equivalent to addition modulo $2$), defined as $ 0\oplus b = b$, and $ 1\oplus b = \neg{b}$,
where by $\neg{b}$ we denote negation (NOT) of $b$. This explains why the gate is called
the controlled-NOT (CNOT). The first qubit ($|a\rangle$) is a control qubit, while the 
second ($|b\rangle$) is a target qubit. The first qubits acts as a switch, which turns on negation
of the second quibit if $a=1$ and remain the second qubit unchanged if $a=0$.  

The diagrammatic representation of the of the unitary operator $\hat{U}$ composed of 
elementary quantum gates is called a \emph{quantum circuit}, examples of which can 
be found through this article. Each computational qubit is associated with a horizontal 
line, which arranges the order at which the operations are performed (direction of time). 
The operations are executed from the left to the right. Then, the symbols representing 
gates can be place on either a single-qubit line (e.g. X,Y,Z,H gates) or by joining two or 
more lines (e.g. CNOT, Toffoli gates). 
 
One of the advantages of quantum algorithms is the possibility of implementation the so-called 
\emph{quantum parallelism}, which allows to reduce computational complexity of certain problems. 
The most known example is the Shor algorithm \cite{Shor} which allow to reduce classical NP 
complexity of the factorization problem into the BQP complexity class (see e.g. Ref. \cite{Zoo} 
for definitions of complexity classes). Another seminal example is the Grover algorithm \cite{Grover:1996rk} 
which, statistically, reduces the number of steps needed to find an element in the random database 
containing $N$ elements from classical $N/2$ to $\mathcal{O}(\sqrt{N})$. Even if we do not make 
use of quantum parallelism and the resulting reduction of computational complexity in this paper, 
the methods may also find application in the context of simulations of spin networks. This especially 
concerns the Quantum Phase Estimation Algorithm \cite{Cleve:1997dh} which may possibly be 
applied to effectively measure phases of the spin foam amplitudes.


\begin{thebibliography}{99}

\bibitem{Ambjorn:2012jv}
  J.~Ambjorn, A.~Goerlich, J.~Jurkiewicz and R.~Loll,
  ``Nonperturbative Quantum Gravity,''
  Phys.\ Rept.\  {\bf 519} (2012) 127
  [arXiv:1203.3591 [hep-th]].
  
\bibitem{Ambjorn:2011cg}
  J.~Ambjorn, S.~Jordan, J.~Jurkiewicz and R.~Loll,
  ``A Second-order phase transition in CDT,''
  Phys.\ Rev.\ Lett.\  {\bf 107} (2011) 211303
  [arXiv:1108.3932 [hep-th]].
  
\bibitem{Bilke:1994yf}
  S.~Bilke, Z.~Burda and J.~Jurkiewicz,
  ``Simplicial quantum gravity on a computer,''
  Comput.\ Phys.\ Commun.\  {\bf 85} (1995) 278
  [hep-lat/9403017].
  
\bibitem{DiFrancesco:1993cyw}
  P.~Di Francesco, P.~H.~Ginsparg and J.~Zinn-Justin,
  ``2-D Gravity and random matrices,''
  Phys.\ Rept.\  {\bf 254} (1995) 1
  [hep-th/9306153].
  
\bibitem{tHooft:1973alw}
  G.~'t Hooft,
  ``A Planar Diagram Theory for Strong Interactions,''
  Nucl.\ Phys.\ B {\bf 72} (1974) 461.
  
\bibitem{Ashtekar:2004eh}
  A.~Ashtekar and J.~Lewandowski,
  ``Background independent quantum gravity: A Status report,''
  Class.\ Quant.\ Grav.\  {\bf 21} (2004) R53.  

\bibitem{CRFV}
C.~Rovelli, F.~Vidotto, ``Covariant Loop Quantum Gravity: An Elementary 
Introduction to Quantum Gravity and Spinfoam Theory,'' Cambridge Monographs 
on Mathematical Physics, 2014.   

\bibitem{Rovelli:1995ac}
  C.~Rovelli and L.~Smolin,
  ``Spin networks and quantum gravity,''
  Phys.\ Rev.\ D {\bf 52} (1995) 5743
  [gr-qc/9505006].

\bibitem{Baez:1997zt}
  J.~C.~Baez,
  ``Spin foam models,''
  Class.\ Quant.\ Grav.\  {\bf 15} (1998) 1827
  [gr-qc/9709052].
  
\bibitem{Perez:2012wv}
  A.~Perez,
  ``The Spin Foam Approach to Quantum Gravity,''
  Living Rev.\ Rel.\  {\bf 16} (2013) 3
  [arXiv:1205.2019 [gr-qc]].
  
\bibitem{Engle:2007uq}
  J.~Engle, R.~Pereira and C.~Rovelli,
  ``The Loop-quantum-gravity vertex-amplitude,''
  Phys.\ Rev.\ Lett.\  {\bf 99} (2007) 161301
  [arXiv:0705.2388 [gr-qc]].
  
\bibitem{Bianchi:2012nk}
  E.~Bianchi and F.~Hellmann,
  ``The Construction of Spin Foam Vertex Amplitudes,''
  SIGMA {\bf 9} (2013) 008
  [arXiv:1207.4596 [gr-qc]].
  
\bibitem{Oriti:2006se}
  D.~Oriti,
  In *Oriti, D. (ed.): Approaches to quantum gravity* 310-331
  [gr-qc/0607032].
  
\bibitem{Freidel:2005qe}
  L.~Freidel,
  ``Group field theory: An Overview,''
  Int.\ J.\ Theor.\ Phys.\  {\bf 44} (2005) 1769
  [hep-th/0505016].
  
\bibitem{Krajewski:2012aw}
  T.~Krajewski,
  ``Group field theories,''
  PoS QGQGS {\bf 2011} (2011) 005
  [arXiv:1210.6257 [gr-qc]].
  
\bibitem{Ooguri:1992eb}
  H.~Ooguri,
  ``Topological lattice models in four-dimensions,''
  Mod.\ Phys.\ Lett.\ A {\bf 7} (1992) 2799
  [hep-th/9205090].
  
\bibitem{Gielen:2013kla}
  S.~Gielen, D.~Oriti and L.~Sindoni,
  ``Cosmology from Group Field Theory Formalism for Quantum Gravity,''
  Phys.\ Rev.\ Lett.\  {\bf 111} (2013) no.3,  031301
  [arXiv:1303.3576 [gr-qc]].
  
\bibitem{Benedetti:2014qsa}
  D.~Benedetti, J.~Ben Geloun and D.~Oriti,
  ``Functional Renormalisation Group Approach for Tensorial Group Field Theory: a Rank-3 Model,''
  JHEP {\bf 1503} (2015) 084
  [arXiv:1411.3180 [hep-th]].
  
\bibitem{Li:2017gvt}
  K.~Li {\it et al.},
  ``Quantum Spacetime on a Quantum Simulator,''
  arXiv:1712.08711 [quant-ph].
 
\bibitem{Feller:2015yta}
A.~Feller and E.~R.~Livine,
``Ising Spin Network States for Loop Quantum Gravity: a Toy Model for Phase Transitions,''
Class.\ Quant.\ Grav.\  {\bf 33} (2016) no.6,  065005.      

\bibitem{Mielczarek:2018ttq}
J.~Mielczarek,
``Spin networks on adiabatic quantum computer,''
arXiv:1801.06017 [gr-qc].  

\bibitem{IBM}
https://www.research.ibm.com/ibm-q/

\bibitem{You} 
J.~Q.~You, F.~Nori, 
``Superconducting Circuits and Quantum Information,'' Phys. Today {\bf 58} (11), 42 (2005).  

\bibitem{QX}
http://quantum-studio.net/

\bibitem{QuantInsp}
https://www.quantum-inspire.com/ 
  
\bibitem{Feynman:1981tf}
R.~P.~Feynman,
``Simulating physics with computers,''
Int.\ J.\ Theor.\ Phys.\  {\bf 21} (1982) 467.  

\bibitem{Mielczarek:2018nnd}
  J.~Mielczarek,
  ``Quantum Gravity on a Quantum Chip,''
  arXiv:1803.10592 [gr-qc].
  
\bibitem{AQG}  
T. ~Albash, D. ~A.~Lidar,  ``Adiabatic Quantum Computing,'' Rev. Mod. Phys. {\bf 90}, 015002 (2018).

\bibitem{Deutsch:1995dw}
  D.~Deutsch, A.~Barenco and A.~Ekert,
  ``Universality in quantum computation,''
  Proc.\ Roy.\ Soc.\ Lond.\ A {\bf 449} (1995) 669
  [quant-ph/9505018].

\bibitem{Ekert}
A.~Ekert, P.~Hayden, H.~Inamori, ``Basic concepts in quantum computation,'' arXiv:quant-ph/0011013.

\bibitem{Chen}
64-Qubit Quantum Circuit Simulation
Zhao-Yun~Chen, et al. ``64-Qubit Quantum Circuit Simulation,'' Science Bulletin, 
2018, {\bf 63}(15):964-971, [arXiv:1802.06952]. 

\bibitem{Biamonte}
J.~D.~Biamonte, M.~E.~S.~Morales, D.~E~Koh, ``Quantum Supremacy Lower Bounds by 
Entanglement Scaling,'' arXiv:1808.00460.

\bibitem{Long}
Gui-Lu Long, Yang Sun, ``Efficient Scheme for Initializing a Quantum Register with an 
Arbitrary Superposed State,'' Phys. Rev. A {\bf 64} (2001) 014303, [arXiv:quant-ph/0104030v1]. 

\bibitem{Mielczarek2019}
J.~Mielczarek, in preparation (2019). 

\bibitem{Quirk} 
https://algassert.com/quirk

\bibitem{QKit}
https://sites.google.com/view/quantum-kit/home

\bibitem{Nielsen}
M.~A.~Nielsen,  I.~L.~Chuang, ``Quantum Computation and Quantum Information,'' 
Cambridge University Press, Cambridge, UK, 2000.

\bibitem{Oeckl:2003vu}
  R.~Oeckl,
  ``A 'General boundary' formulation for quantum mechanics and quantum gravity,''
  Phys.\ Lett.\ B {\bf 575} (2003) 318
  [hep-th/0306025].

\bibitem{Baytas:2018wjd}
  B.~Baytas, E.~Bianchi and N.~Yokomizo,
  ``Gluing polyhedra with entanglement in loop quantum gravity,''
  Phys.\ Rev.\ D {\bf 98} (2018) no.2,  026001
  [arXiv:1805.05856 [gr-qc]].
  
\bibitem{Maldacena:1997re}
  J.~M.~Maldacena,
  ``The Large N limit of superconformal field theories and supergravity,''
  Int.\ J.\ Theor.\ Phys.\  {\bf 38} (1999) 1113
   [Adv.\ Theor.\ Math.\ Phys.\  {\bf 2} (1998) 231]
  [hep-th/9711200].
  
\bibitem{Susskind:2017ney}
  L.~Susskind,
  arXiv:1708.03040 [hep-th].
  
\bibitem{Swingle:2009bg}
  B.~Swingle,
  ``Entanglement Renormalization and Holography,''
  Phys.\ Rev.\ D {\bf 86} (2012) 065007
  doi:10.1103/PhysRevD.86.065007
  [arXiv:0905.1317 [cond-mat.str-el]].
 
\bibitem{Ryu:2006bv}
  S.~Ryu and T.~Takayanagi,
  ``Holographic derivation of entanglement entropy from AdS/CFT,''
  Phys.\ Rev.\ Lett.\  {\bf 96} (2006) 181602
  [hep-th/0603001].
  
\bibitem{Rangamani:2016dms}
  M.~Rangamani and T.~Takayanagi,
  ``Holographic Entanglement Entropy,''
  Lect.\ Notes Phys.\  {\bf 931} (2017) pp.1
  [arXiv:1609.01287 [hep-th]].
    
\bibitem{Han:2016xmb}
  M.~Han and L.~Y.~Hung,
  ``Loop Quantum Gravity, Exact Holographic Mapping, and Holographic Entanglement Entropy,''
  Phys.\ Rev.\ D {\bf 95} (2017) no.2,  024011
  [arXiv:1610.02134 [hep-th]].
  
\bibitem{tensornetworks}
J.~Biamonte, V.~Bergholm, ``Tensor Networks in a Nutshell,'' [arXiv:1708.00006].

\bibitem{Bojowald:2008zzb}
  M.~Bojowald,
  ``Loop quantum cosmology,''
  Living Rev.\ Rel.\  {\bf 11} (2008) 4.

\bibitem{Bianchi:2010zs}
  E.~Bianchi, C.~Rovelli and F.~Vidotto,
  ``Towards Spinfoam Cosmology,''
  Phys.\ Rev.\ D {\bf 82} (2010) 084035
  [arXiv:1003.3483 [gr-qc]].

\bibitem{Vidotto:2010kw}
  F.~Vidotto,
  ``Spinfoam Cosmology: quantum cosmology from the full theory,''
  J.\ Phys.\ Conf.\ Ser.\  {\bf 314} (2011) 012049
  [arXiv:1011.4705 [gr-qc]].

\bibitem{Czech:2018kvg}
  B.~Czech, L.~Lamprou and L.~Susskind,
  ``Entanglement Holonomies,''
  arXiv:1807.04276 [hep-th].
  
\bibitem{Dona:2017dvf}
  P.~Donà, M.~Fanizza, G.~Sarno and S.~Speziale,
  Class.\ Quant.\ Grav.\  {\bf 35} (2018) no.4,  045011
  [arXiv:1708.01727 [gr-qc]].
  
\bibitem{Sarno:2018ses}
  G.~Sarno, S.~Speziale and G.~V.~Stagno,
  Gen.\ Rel.\ Grav.\  {\bf 50} (2018) no.4,  43
  [arXiv:1801.03771 [gr-qc]].
  
\bibitem{Dona:2018nev}
  P.~Dona and G.~Sarno,
  Gen.\ Rel.\ Grav.\  {\bf 50} (2018) 127
  [arXiv:1807.03066 [gr-qc]].
  
\bibitem{Dona:2019dkf}
  P.~Dona, M.~Fanizza, G.~Sarno and S.~Speziale,
  arXiv:1903.12624 [gr-qc].
  
\bibitem{Oriti:2017twl}
  D.~Oriti,
  ``Spacetime as a quantum many-body system,''
  arXiv:1710.02807 [gr-qc].
  
\bibitem{Cleve:1997dh}
  R.~Cleve, A.~Ekert, C.~Macchiavello and M.~Mosca,
  ``Quantum algorithms revisited,''
  Proc.\ Roy.\ Soc.\ Lond.\ A {\bf 454} (1998) 339
  [quant-ph/9708016].

\bibitem{Kadowaki}  
T.~Kadowaki, H.~Nishimori, ``Quantum annealing in the transverse Ising model,'' 
Phys. Rev. E {\bf 58}, 5355 (1998). 

\bibitem{DWave}  
https://www.dwavesys.com/home   

\bibitem{TangleLake} 
https://newsroom.intel.com/press-kits/quantum-computing/\#quantum-computing

\bibitem{QuTech} 
https://qutech.nl/

\bibitem{Google} 
https://ai.google/research/teams/applied-science/quantum-ai/

\bibitem{Rigetti} 
https://www.rigetti.com/ 

\bibitem{Microsoft} 
https://www.microsoft.com/en-us/quantum/

\bibitem{Ionq} 
https://ionq.co/

\bibitem{Report}
Ed. E.~Grumbling and M.~Horowitz, ``Quantum Computing - Porgress and Prospects,''
The National Academies Press, 2019.  

\bibitem{QuantVol}
A.~W.~Cross  et al., ``Validating quantum computers using randomized model circuits,'' arXiv:1811.12926 [quant-ph].  

\bibitem{IBMQuantVol}
https://www.ibm.com/blogs/research/2019/03/power-quantum-device/

\bibitem{Moore}
G.~E.~Moore, ``Cramming more components onto integrated circuits,'' Electronics, Vol. 38, 
No. 8, 1965. 

\bibitem{Shor}
P. W. Shor, "Algorithms for quantum computation: discrete logarithms and factoring," Proceedings 35th 
Annual Symposium on Foundations of Computer Science, Santa Fe, NM, USA, 1994, pp. 124-134.

\bibitem{Zoo}
https://complexityzoo.uwaterloo.ca/Complexity\_Zoo  

\bibitem{Grover:1996rk}
  L.~K.~Grover,
  ``A Fast quantum mechanical algorithm for database search,''
  Proceedings, 28th Annual ACM Symposium on the Theory of Computing, 1996, 
 [quant-ph/9605043].

  
\end{thebibliography}
\end{document}